%% file: new_skew.tex
\documentclass[32pt, usenatbib]{mn2e}
 \usepackage{graphicx,times,subfigure}
\usepackage{bm}
\usepackage{epsf}
\usepackage{amsmath}
\usepackage{amssymb}
\usepackage{cases}
\usepackage{float}
\usepackage{colortbl}
\usepackage{color}
\usepackage{multirow}
\usepackage{url}
\usepackage{hyperref}

\def\bkp{{\bf k}^{\prime}}
\def\rp{r^{\prime}}

\usepackage{xcolor}
\def\xred{\textcolor{black}}

\input{local-macro}

\begin{document}
\onecolumn
\title[Higher-Order Spectra of Weak Lensing Convergence Maps]
      {Higher-Order Spectra of Weak Lensing Convergence Maps\\ in Parameterized Theories of Modified Gravity}
      \author[Munshi, McEwen]
             {D. Munshi, J. D. McEwen \\
               Mullard Space Science Laboratory, University College London,
               Holmbury St Mary, Dorking, Surrey RH5 6NT, UK}
\maketitle
\begin{abstract}
{We compute the low-$\ell$ limit of the family of higher-order spectra for projected (2D)
weak lensing convergence maps. In this limit these
spectra are computed to an arbitrary order using {\em tree-level} perturbative calculations.
We use the flat-sky approximation and Eulerian
perturbative results based on a generating function approach.
We test these results for the lower-order members of this family, i.e. the skew- and kurt-spectra
against state-of-the-art simulated all-sky weak lensing convergence maps and find our results to be in
very good agreement. We also show how these spectra can be computed in the presence of a realistic
sky-mask and Gaussian noise. We generalize these results to three-dimensions (3D) and compute the
{\em equal-time} higher-order spectra.
These results will be valuable in analyzing higher-order statistics
from future all-sky weak lensing surveys such as the {\em Euclid} survey at low-$\ell$  modes.
As illustrative examples, we compute these statistics 
in the context of the {\em Horndeski} and {\em Beyond Horndeski} theories of modified gravity.
They will be especially useful in constraining theories such as the Gleyzes-Langlois-Piazza-Vernizzi (GLPV) theories and
Degenerate Higher-Order Scalar-Tensor (DHOST) theories
as well as the commonly used normal-branch of
Dvali-Gabadadze-Porrati (nDGP) model, clustering quintessence models and scenarios with massive neutrinos.}
\end{abstract}
\section{\bf Introduction}
\label{sec:intro}
%
%
We have a standard model of cosmology thanks to recently completed Cosmic Microwave Background Radiation (CMBR) experiments
such as the {\it Planck} Surveyor\footnote{\href{http://http://sci.esa.int/planck/}{\tt Planck}}\citep{Planck1,Planck2,Planck18}.
However, many of the outstanding questions pertaining , e.g. to the nature of dark matter
and dark energy or possible modification of
General Relativity on cosmological scales remain unanswered \citep{MG1,MG2}. In addition they will also
provide an estimate of the sum of the neutrino masses \citep{nu}. Ongoing and planned
future large scale structure (LSS) surveys
may resolve or will provide clues for these questions using weak lensing analyses.
Observational programs of many such surveys, including
{\it Euclid}\footnote{\href{http://sci.esa.int/euclid/}{\tt http://sci.esa.int/euclid/}}\citep{Euclid}, 
{CFHTLS}\footnote{\href{http://www.cfht.hawai.edu/Sciences/CFHLS/}{\tt http://www.cfht.hawai.edu/Sciences/CFHLS}},
{PAN-STARRS}\footnote{\href{http://pan-starrs.ifa.hawai.edu/}{\tt http://pan-starrs.ifa.hawai.edu/}},
Dark Energy Surveys (DES)\footnote{\href{https://www.darkenergysurvey.org/}{\tt https://www.darkenergysurvey.org/}}\citep{DES},
{WiggleZ}\footnote{\href{http://wigglez.swin.edu.au/}{\tt http://wigglez.swin.edu.au/}}\citep{WiggleZ},
{LSST}\footnote{\href{http://www.lsst.org/llst home.shtml}{\tt {http://www.lsst.org/llst home.shtml}}}\citep{LSST_Tyson},
{BOSS}\footnote{\href{http://www.sdss3.org/surveys/boss.php}{\tt http://www.sdss3.org/surveys/boss.php}}\citep{SDSSIII},
{KiDS}\citep{KIDS} and WFIRST\citep{WFIRST} lists weak lensing as their main science driver.

From the early days of detection the weak lensing \citep{review} studies have 
matured to a point where weak lensing results from {\it Euclid} are expected to constrain the cosmological
parameters to sub-percent accuracy.
However, weak lensing at smaller angular scales probes the nonlinear regime of gravitational
clustering, and is thus key to understanding the non-Gaussianity induced by the
nonlinearity and fullly exploiting in the weak lensing maps.
The higher-order statistics are also useful for the breaking of parameter degeneracies in
studies involving the power spectrum analysis alone and they are also important in
understanding the variance or error of estimation of lower-order statistics.
%
%
These higher-order statistics
including the cumulants \citep{Bernard1} and their correlators \citep{Bernardeau_bias,joint,Spergel,Erminia} are
among the best-known diagnostics of the deviation from Gaussianity that
characterizes the non-linear regime \citep{review_ng}, with a long history
analytical modeling \citep{bernardeau_review}.
Most of
these studies use extensions of perturbative results in the
quasilinear regime valid at large smoothing angular scales or
variants of halo models \citep{CooraySheth}. Early studies concentrated on measurements of
higher-order correlation hierarchy in the angular space due to small survey size \citep{Mellier1,Mellier2}.
However, the near all-sky coverage of future surveys such as {\it Euclid} 
will let us estimate higher-order statistics in the harmonic domain with
unprecedented accuracy \citep{Euclid_Review}.
While measurements of real space correlations are much simpler in
the presence of complicated survey design the measurements for different
angular scales can be highly correlated \citep{MunshiJain2,Munshi_bias}. In comparison measurements
in the harmonic domain are less correlated and each mode contains (nearly) independent information
in the limit of all- sky coverage.
The primary motivation of this study is to develop 
analytical predictions for one such statistic called the skew-spectrum,
and test them against numerical simulations.
We will borrow the concepts developed for constructing skew-spectra for the the study of
non-Gaussianity in the context of CMBR observations \citep{Planck}. However, we also include
gravity-induced secondary non-Gaussianity.
The skew-spectrum is the lowest-order member in the family of higher-order spectra \citep{xn1,xn2}.
%
%
In a series of papers the one-point statistics such as the skewness and
kurtosis were generalized to two-point cumulant correlator, e.g.
the two-to-one correlator and its higher-order generalizations.
These can be represented in the harmonic domain by their associated
spectra such as the skew-spectrum \citep{MunshiHeavens} and its higher-order generalizations \citep{xn1,xn2}.
These spectra have already been used to analyze
WMAP\footnote{\href{https://map.gsfc.nasa.gov/}{\tt https://map.gsfc.nasa.gov/}}\citep{Smidt1}
as well as {\it Planck} data \citep{Planck}. They are useful
tools to separate individual contributions and estimate
systematics. In this paper we will concentrate on the projected
skew-spectrum and kurt-spectrum in the context of weak lensing surveys \citep{secondary2}.

Other similar estimators also exist, including the morphological estimators
\citep{waerbeke}, e.g. position-dependent power spectra \citep{IB_MG},
phase correlations \citep{phase_statistics}, line-correlations \citep{line_corr},
peak-statistics \citep{peak_count},
peak-correlations \citep{peak_correlation} and extreme value statistics \citep{extreme}.

Many modified gravity theories are now severely constrained with the first detection of GW170817  \citep{GW1}
and its electromagnetic counterpart GRB 170817A \citep{Goldstein} implying Gravity Waves travel at the speed of light
with deviation smaller than $\rm few \times 10^{-15}$
- see e.g. \cite{Baker17, Jain17, Lucas17, CremVer17}. However, some of the models we consider here
are designed to evade this constraint.
It is expected that the constraints on these models will be further tightened by the observations
of large scale structure by {\it Euclid} and LSST. The higher-order statistics
we develop here can be very effectively used to test these models independently or jointly with power spectrum
estimates. As a concrete example of the higher-order spectra we take the modified gravity
theories also known as the Horndeski's theory of gravity. These are the most general
theory of gravity that has second-order equation of motion.
It was proposed first in 1974 \citep{hordenski74} and since then, it was realised
that Horndeski theory contains many other theories of gravity as a special cases.
These include General relativity, Jordon-Brans-Dicke theories of gravity, Dilaton and
Chameleon theories of gravity, theories involving as co-variant Galileons as well as
models of Quintessence. All of these models of gravity have found use
in construction of cosmological models of inflation as well as dark energy
(see e.g. \cite{Deff11, kobayeshi11, BeyondHordensky1,BeyondHordensky2,LN16a,LN16b} for an incomplete list
of recent references). We use a recent parametrization of the gravity induced bispectrum in this model
as well as models that are known as the beyond Horndeski theories to compute
the skew-spectrum in the low-$\ell$ limit.

This paper is organized as follows.
In \textsection\ref{sec:bispec} we review results regarding the
convergence bispectrum in the context of tree-level Standard Perturbation Theory (SPT).
In \textsection\ref{sec:skew}
we introduce the skew-spectrum and relate it to the bispectrum.
The corresponding results for trispectrum and kurt-spectra are derived in \textsection\ref{sec:kurt}.
Theoretical predictions in the context of generating functions are derived in \textsection{\ref{sec:tophat}},
The generalization of higher-order spectra is presented in \textsection{\ref{sec:3D}}.
The higher-order spectra can be derived in the presence of a mask.
The corresponding results are presented in \textsection{\ref{sec:pcl}},
The simulations are discussed in \textsection{\ref{sec:simu}},
the numerical results are presented in \textsection{\ref{sec:num}}.
We present the results for various modified gravity and other beyond-$\Lambda$CDM scenarios in \textsection\ref{sec:modG}.
Finally, the conclusions are drawn in \textsection{\ref{sec:conclu}}.

%
\section{Modelling of Higher-order Weak Lensing Spectra}
\label{sec:bispec}
In this section we will review the aspects of standard tree-level perturbative
which we use to compute the bispectrum as well trispectrum as and eventually
the skew-spectrum and kurt-spectrum.
\subsection{Tree-level Perturbative Calculations}
\label{subsec:tree}
%
In the quasilinear regime ($\delta \le 1$), the evolution of
density contrast $\delta$ can be described using SPT
\citep{review}. However, the treatment based on perturbation theory breaks
down when density contrast at a given length scale becomes nonlinear
($\delta \ge 1$) which significantly increases the growth of
clustering. We will denote the Fourier transform of the
density contrast $\delta({\bf r})$ by $\delta(\bk)$,
where ${\bf r}$ is the comoving co-ordinate, and ${\bf k}$
denotes the comoving wavenumber.
Expanding the $\delta({\bf k})$, in a perturbative
series, and assuming the density contrast is less than unity, for
the pertubative series to be convergent, we get:
\bes\ben
&&\delta({\bf k}) = \delta^{(1)}({\bf k}) + \delta^{(2)}({\bf k})
+ \delta^{(3)}({\bf k}) + \dots, \label{eq:pert}.
\label{eq:define_kappa}
\een
The $n$-th order perturbative term denoted as $\delta^{(n)}$ 
is $\propto [\delta^{(1)}]^n$ where $\delta^{(1)}$ is the linear density contrast.
The term  $\delta^{(n)}$ is expressed using a kernel $F_n$ using the following convolution:
\ben
&&\delta^{(n)}(\bk) = \int d\bk_1.
\cdots
\int d\bk_n
\delta_{\rm 3D}({\bf k_1 + \cdots  +
k_n -k }) F_n(\bk_1, \cdots ,\bk_n) \delta^{(1)}({\bf k}_1)\cdots \delta^{(1)}({\bf
  k}_n); \quad d\bk= {d^3\bk \over (2\pi)^{3/2}} \label{eq:F2}.
\een
The Dirac delta function in $\rm 3D$ is denoted by $\delta_{\rm 3D}$ and $\bk_1,\bk_2,\cdots,\bk_n$
denotes different wavenumbers. The second-order kernel $F_2$ has the following expression.
For the higher-order kernels see \citep{review}:
\ben
&& F_2(\bk_1,\bk_2) = {5\over 7} +{1\over 2}\left({k_1\over k_2}+{k_2\over k_1}\right ) 
\left({\bk_1\cdot\bk_2 \over k_1 k_2}\right )
+{2 \over 7}
\left( {\bk_1\cdot\bk_2 \over k_1 k_2}\right )^2, \quad k_i = |\bk_i|.
\label{eq:F3}
\een\ees
Throughout we will use the following convention for the three-dimensional (3D) Fourier Transform (FT) and its inverse:
\ben
&& \delta({\bf k}) = \int d{\bf r} \exp(-i {\bf k} \cdot{\bf r}) \delta({\bf r}); \quad
\delta({\bf r}) =  \int d{\bf k} \exp(i\,{\bf k}\cdot{\bf r}) \delta({\bf k}); \quad\quad
d\br= {d^3\br \over (2\pi)^{3/2}}.
\label{eq:Fourier_cartesian}
\een
We have suppressed the temporal dependence in Eq.(\ref{eq:pert})-Eq.(\ref{eq:F2})
which will be introduced later in this section.
The power spectrum $P_{\delta}$ and the bispectrum $B_{\delta}$ of $\delta$ are defined respectively
as the two and three point correlation of the variable $\delta({\bf k})$
The $P_{\rm lin}(k)$ denotes the linear power spectrum, i.e.
$\delta_{\rm lin}({\bf k}) = \delta^{(1)}({\bf k})$ and
$\la\delta_{\rm lin}(\bk_1)\delta_{\rm lin}(\bk_2)\ra_c= (2\pi)^3\delta_{\rm 3D}(\bk_1+\bk_2)P_{\rm lin}(k_1)$.
Throughout angular brackets represent ensemble averaging. The subscript ${}_{\rm lin}$
stands for linear-order contributions.

The linearized solution for the density field is
$\delta^{(1)}({\bf k})$; higher-order terms yield corrections to
this linear solution. Using an {\em ideal} fluid approach known to be valid at
large scales (and before shell crossing) one can write the second
order correction to the linearized density field using the kernel
$F_2({\bf k_1},{\bf k_2})$. The cosmological structure formation is described by a
set of equation which describes the Newtonian gravity coupled to the Euler
and continuity equation \citep{review}.  This system of non-linear, 
coupled integro-differential equations are used to compute the kernels
$F_2(\bk_1,\bk_2)$, $F_3(\bk_1,\bk_2,\bk_3)$ and their high-order counterparts.
This is typically done perturbatively in an order-by-order manner.
%
\subsection{Weak Lensing Statistics in Projection (2D)}
\label{sec:flat_sky}
%
We will now specialize our discussion to weak lensing surveys. The weak lensing convergence $\kappa$ is a
line of sight projection of the 3D density contrast $\delta({\bf r})$:
\ben
&& \kappa({\oh}; r_s) = \int_0^{r_s}\, dr\, w(r, r_s)\, \delta(r,{\oh}); \quad 
w(r, r_s) = {3 \Omega_{\rm M} \over 2}\, {H_0^2 \over a c^2}\, \, {d_A(r) d_{A}({r_s-r}) \over d_A(r_s)}. \label{eq:define_omega}
\een
Where $r$ is the comoving distance, $\oh=(\theta,\phi)$ is a unit vector that defines
the position of the ppixel on the surface of the sky, with $\theta$ and $\phi$ respectively representing the
azimuthal and polar co-ordinates $d\oh =\sin\theta\, d\theta\, d\varphi$ is the measure of integration,
$r_s$ is the radial comoving distance to
the source plane, $c$ is the speed of light, $a$ represents the scale factor,
$H_0$ the Hubble parameter, $d_A(r)$ is the {comoving} angular diameter distance and the
three-dimensional (3D) density contrast $\delta$ and $\Omega_{\rm M}$ is the cosmological density parameter.
We will ignore the source distribution and assume them to be localized on a single source plane, we will also ignore photometric redshift errors.
However, such complications are essential to link predictions to observational data and can readily be included in our analysis.
To avoid cluttering, we will suppress the $r_s$ dependence of $\kappa(\oh, r_s)$ and $w(r, r_s)$ defined in Eq.(\ref{eq:define_kappa}) in the following.
The corresponding 3D power spectrum $P_{\delta}$, bispectrum $B_{\delta}$ and trispectrum $T_{\delta}$ for $\delta$ are:
\bes
\ben
&& \la\delta({\bk}_{1})\delta({\bk}_{2})\ra_c =
(2\pi)^{3}\delta_{\rm 3D}({\bk}_{1}+{\bk}_{2})P_{\delta}(k_{1}); \quad k=|{\bf k}|;\\
&& \la\delta({\bk}_{1})\delta({\bk}_{2})\delta({\bk}_{3})\ra_c 
= (2\pi)^{3}\delta_{\rm 3D}({\bk}_{1}+{\bk}_{2}+{\bk}_{3})B_{\delta}({\bk}_{1},{\bk}_{2},{\bk}_{3}); \\
&& \la\delta({\bk}_{1})\cdots \delta({\bk}_{4})\ra_c 
= (2\pi)^{3}\delta_{\rm 3D}({\bk}_{1}+\cdots+{\bk}_{4})T_{\delta}({\bk}_{1},\cdots,{\bk}_{3}).
\een
\ees
The subscript ${}_c$ denotes the fact that only connected diagrams are included in computing these statistics.
The flat-sky power spectrum $P^{\kappa}$ and bispectrum $B^{\kappa}$ are similarly defined through \citep{review}:
\bes
\ben
&& \langle \kappa(\bl_1)\kappa(\bl_2)\rangle_c = 
(2\pi)^2 \delta_{\rm 2D}(\bl_1+\bl_2) P^\kappa(l_1); \\
&& \langle \kappa(\bl_1)\kappa(\bl_2)\kappa(\bl_3)\rangle_c =
(2\pi)^2 \delta_{\rm 2D}(\bl_1+\bl_2+\bl_3) B^\kappa(\bl_1,\bl_2,\bl_3); \\
&& \langle \kappa(\bl_1) \cdots \kappa(\bl_4)\rangle_c =
(2\pi)^2 \delta_{\rm 2D}(\bl_1+\cdots+\bl_3) T^\kappa(\bl_1,\bl_2,\bl_3,\bl_4). 
\een
\ees
The wavenumbers $\bl,\bl_1, \cdots \bl_4$ are wavenumbers defined on the flat-patch of the sky.
For a given radial distance $r$ they are related to the projected 3D wave number by the relation $\bl = \bk_{\perp}/d_A(r)$;
where $d_{A}(r)$ being the co-moving angular diameter distance defined before and $l=|{\bf l}|$.
Using the {\em flat-sky} approximation as well as {\em Limber} and {\em prefactor unity} approximation
the projected power spectrum $P^{\kappa}(\bl)$ and bispectrum $B^{\kappa}({\bf k}_1,{\bf k}_2,{\bf k}_3)$ can be expressed
respectively in terms of the 3D $\delta$ power spectrum $P_{\delta}(k)$ and bispectrum  $B_{\delta}({\bf k}_1,{\bf k}_2,{\bf k}_3)$ \citep{MunshiReview}:
\bes
\ben
&& P^{\kappa}({l}) = \int_0^{r_s} dr {\omega^2(r) \over d_A^2(r)}
P_{\delta}\left ({{l} \over d_A(r)}; r \right ); \label{eq:weak_ps}\\
&& B^{\kappa}({\bl}_{1},{\bl}_{2},{\bl}_{3}) = \int_0^{r_s} dr {\omega^3(r) \over d_A^4(r)}
B_{\delta}\left ({{\bl}_{1} \over d_A(r)},{{\bl}_{2} \over d_A(r),},
{{\bl}_{3} \over d_A(r)}; r \right); \label{eq:weak_bps}\\
&& T^{\kappa}({\bl}_{1},{\bl}_{2},{\bl}_{3},{\bl}_{4}) = \int_0^{r_s} dr {\omega^4(r) \over d_A^6(r)}
T_{\delta}\left ({{\bl}_{1} \over d_A(r)},{{\bl}_{2} \over d_A(r),},
{{\bl}_{3} \over d_A(r)}, {{\bl}_{4} \over d_A(r)} ; r \right). \label{eq:weak_tps}
\een
\ees
The superscript $\kappa$ correspond to the convergence field which these statistics correspond to.
The function $\omega$ is defined in Eq.(\ref{eq:define_omega}).
We will use different approximations introduced
in \textsection\ref{sec:bispec}  in Eq.(\ref{eq:weak_ps})-Eq.(\ref{eq:weak_bps}) to compute
the convergence or $\kappa$ bispectrum.

\section{Bispectrum and Skew-spectrum}
\label{sec:skew}
%
The spherical harmonic transform of a convergence map $\kappa(\oh)$, denoted as $\kappa_{\ell m}$,
defined over the surface of the sky using spherical harmonics $Y_{\ell m}(\oh)$
can be used to define the multipoles $\kappa_{\ell m}$:
\ben
&& \kappa_{\ell m} = \int\, d{\oh}\, Y_{\ell m}(\oh)\, \kappa({\oh});
\quad \oh=(\theta,\varphi).
\een
A Gaussian field is completely characterized by its 
power spectrum ${\cal C}^{\kappa}_{\ell}$ which is defined as
${\cal C}^{\kappa}_{\ell} = \langle \kappa_{\ell m}\kappa^*_{\ell m}\rangle$.
Here $\kappa^*_{\ell m}$ represents the complex conjugate of $\kappa_{\ell m}$.
The flat sky power spectrum $P^{\kappa}(l)$ is identical
to ${\cal C}^{\kappa}_{\ell}$ at high $\ell$ with the identification $l=\ell$.
Bispectrum is the lowest order statistics that characterizes departure from
Gaussianity that is defined as the three-point coupling of
harmonic coefficients.
Assuming isotropy and homogeneity the all-sky bispectrum $B^\kappa_{\ell_1\ell_2\ell_3}$ is defined as \citep{review_ng}:
\ben
&& \la \kappa_{\ell_1 m_1}\kappa_{\ell_2 m_2}\kappa_{\ell_3m_3}\ra_c \equiv 
B^\kappa_{\ell_1\ell_2\ell_3}
\left ( \begin{array} { c c c }
     \ell_1 & \ell_2 & \ell_3 \\
     m_1 & m_2 & m_3
\end{array} \right ).
\een
The quantity in parentheses is the well-known Wigner-3j symbol which
enforces rotational invariance. It is only non-zero for the triplets
$(\ell_1,\ell_2,\ell_3)$ that satisfy the triangular condition and
$\ell_1+\ell_2+\ell_3$ is even.
The reduced bispectrum $b^\kappa_{\ell_1\ell_2\ell_3}$ for convergence $\kappa$
is defined through the following expression \citep{review_ng}:
\ben 
&& B^\kappa_{\ell_1\ell_2\ell_3} = \sqrt{(2\ell_1+1)(2\ell_2+1)(2\ell_3+1)\over 4\pi}
\left ( \begin{array} { c c c }
     \ell_1 & \ell_2 & \ell_3 \\
     0 & 0 & 0
\end{array} \right ) b^\kappa_{\ell_1\ell_2\ell_3}.
\een
The skew-spectrum is defined as the cross power spectrum
formed by cross-correlating the squared $\kappa^2$ maps against
the original map $\kappa$ \citep{MunshiHeavens}:
\bes
\ben
&& {\cal S}^{(21)}_{\ell} =
    {1\over 2\ell+1}\sum_m {\rm Real}\{ [\kappa^2]_{\ell m}[\kappa]^*_{\ell m} \}
    =\sum_{\ell_1\ell_2} B^\kappa_{\ell_1 \ell_2\ell} J_{\ell_1\ell_2\ell}; \label{eq:defSkew} \\
&& J_{\ell_1\ell_2\ell} = \sqrt {   {(2\ell_1+1)(2\ell_2+1) \over ( 2\ell+1 )}  }
\left ( \begin{array} { c c c }
     \ell_1 & \ell_2 & \ell \\
     0 & 0 & 0
  \end{array} \right ). \label{eq:defJ}
\een
\ees
To avoid cluttering we will not explicitly display smoothing windows in our equations.
The beam-smoothed versions of the expressions can be recovered by using the smoothed harmonics i.e. replacing $k_{\ell m}$ with 
$\kappa_{\ell m}b_{\ell}$ where $b_{\ell}$ is the smoothing beam in the harmonic domian which can be tophat or Gaussian.
In case of Gaussian smoothing the expressions are derived in an order-by-order manner \citep{Bernard1,bias96} 
For a tophat smoothing these expressions are derived using a generating function
to an arbitrary order \citep{Matsubara02}.
The normalized one-point skewness parameter \citep{bernardeau_review} $S_3 ={\la\kappa^3\ra_c/\la\kappa^2\ra_c^2}$
can be recovered from the skew-spectrum by constructing
the beam-smoothed third-order moment $\la\kappa^3\ra_c$ \citep{MunshiHeavens}
\ben
&& \mu_3 = \la\kappa^3\ra_c =   \sum_{\ell}(2\ell+1)S^{(21)}_{\ell} = \sum_{\ell_1\ell_2\ell} J_{\ell_1\ell_2\ell}B^{\kappa}_{\ell_1\ell_2\ell}.
\label{eq:skewness}
\een
The normalized skewness parameter $S_3$ is defined as $S_3 = \mu_3/\mu_2^2$ with $\mu_{\rm N}=\la\kappa^{\rm N}\ra_c$
and  $S_{\rm N} = \mu_{\rm N}/\mu_2^{\rm N-1}$

The real space two-to-one correlation function can be defined in terms
of the skew-spectrum as \citep{MunshiHeavens}:
\ben
&& \xi^{(21)}(\theta_{12}) = \langle\kappa^2(\oh_1)\kappa(\oh_2)\rangle_c =
   {1 \over 4\pi} \sum_{\ell} (2\ell+1) S^{(21)}_{\ell}P_{\ell}(\cos\theta_{12}). 
   \label{eq:corr21}
   \een
   Where $P_{\ell}$ represents the Legendre Polynomial, and
   the angular positions $\oh_1$ and $\oh_2$ are separated by an angle $\theta_{12}$.
   Suitably normalized two-to-one correlator is 
the lowest order of a family of statistics also known as cumulant
correlator\citep{Bernardeau_bias,joint,Spergel, Erminia},
which has also been used in the context of weak-lensing surveys
\citep{flexions,Munshi_bias}.

In our notation $\delta_{\rm 2D}$ is the 2D Dirac delta function.
The flat-sky bispectrum $B^\kappa(\bl_1,\bl_2,\bl_3)$ is identical to the reduced bispectrum
$b_{\ell_1\ell_2\ell_2}$ for high multipole \citep{review_ng}. This can be shown
by noting the following asymptotic relationship.
\ben
&& {\cal G}_{\ell_1m_1,\ell_2m_2,\ell_3m_3} \equiv \int d\oh Y_{\ell_1m_1}(\oh)
Y_{\ell_2m_2}(\oh)Y_{\ell_3m_3}(\oh); \nn \\
&&  = \sqrt {   {(2\ell_1+1)(2\ell_2+1)( 2\ell_3+1 ) \over 4\pi}  }
\left ( \begin{array} { c c c }
     \ell_1 & \ell_2 & \ell_3 \\
     0 & 0 & 0
  \end{array} \right )\left ( \begin{array} { c c c }
     \ell_1 & \ell_2 & \ell_3 \\
     m_1 & m_2 & m_3
\end{array} \right ) \approx (2\pi)^2\delta_{\rm 2D}(\bl_1+\bl_2+\bl_3).
\een

  A few comments about the skew-spectrum are in order. One-point statistics such
  as the skewness parameter have the advantage of having high signal-to-noise.
  However, they lack distinguishing power as all the available
  information in the bispectrum is compressed into a single number. In contrast,
  the skew-spectrum, encodes some information on the shape
  of the spectrum, and in principle can allow us to separate
  the contribution from gravity-induced non-Gaussianity or possible
  source of contamination from systematics. Though 
  primordial non-Gaussianity is highly constrained in the light
  of {\em Planck} data, such contributions can also tested using the skew-spectrum.

  In this paper we consider a direct estimator for the skew-spectrum
  as opposed to the optimal estimator developed in \citep{MunshiHeavens}
  where optimality was achieved by using suitable weights to the harmonics
  that incorporates a match filtering as well as saturates the Cramer-Rao
  limit in the limit of weakly non-Gaussian limit. Indeed, a simple
  Fisher matrix based analysis, however, will non-longer be adequate for
  moderately non-Gaussian weak lensing maps. However, optimality
  is not of crucial importance of analysis for weak lensing maps as
  the secondary non-Gaussianity is expected to be detected with much
  higher signal-to-noise. A simpler direct estimator 
  will thus be useful for studying non-Gaussianity
  in weak-lensing maps.
%
  \section{Trispectrum AND Kurt-Spectra}
  \label{sec:kurt}
%
  The near all-sky weak lensing maps from surveys such as Euclid will also allow determination
  of non-Gaussianity statistics beyond the lowest-order, e.g. the fourth-order correlator or the
  trispectrum. Trispectrum can be useful not only to construct the covariance of the power
  spectrum estimator but also as a consistency check for the lower order estimators.
  In this section we will extend the estimator present above
  for the bispectrum to the case of trispectrum.

  The trispectrum $T^{\ell_1\ell_2}_{\ell_3\ell_4}(L)$ can be defined by the following expressions from the
  four-point correlation function of the spherical harmonics  $\kappa_{\ell m}$ for the convergence
  field $\kappa$ \citep{xn1}: 
\bes
\ben
&& \la \kappa_{\ell_1m_1} \kappa_{\ell_2m_2} \kappa_{\ell_3m_3} \kappa_{\ell_4m_4} \ra_c =
\sum_{LM} (-1)^{M} T^{\ell_1\ell_2}_{\ell_3\ell_4}(L)\left ( \begin{array} { c c c }
     \ell_1 & \ell_2 & L \\
     m_1 & m_2 & M
\end{array} \right )
\left ( \begin{array} { c c c }
     \ell_3 & \ell_4 & L \\
     m_3 & m_4 & -M
\end{array} \right );  \\
&& T^{\ell_1\ell_2}_{\ell_3\ell_4}(L) = (2L+1)\sum_{M} \sum_{m_i}
\left ( \begin{array} { c c c }
     \ell_1 & \ell_2 & L \\
     m_1 & m_2 & M
\end{array} \right )
\left ( \begin{array} { c c c }
     \ell_3 & \ell_4 & L \\
     m_3 & m_4 & -M
\end{array} \right )\la \kappa_{\ell_1m_1} \kappa_{\ell_2m_2} \kappa_{\ell_3m_3} \kappa_{\ell_4m_4} \ra_c.
\een
\ees
Here, $M$ is the magnetic quantum number associated with the azimuthal quantum number $L$.
The Wigner 3j-symbols above ensure that the triangle inequalities imposed by statistical isotropy and homogeneity
of the trispectrum in the harmonic space is represented by a quadrilateral. The harmonics $\ell_1$, $\ell_2$, $\ell_3$ and
$\ell_4$ represent the sides of the quadrilateral and the harmonics $L$ represents one of the diagonal of the quadrilateral
The two kurt-spectra ${\cal K}_{\ell}^{(31)}$ and  ${\cal K}_{\ell}^{(31)}$ are defined as \citep{xn1,xn2}:
\bes
\ben
&& {\cal K}_{\ell}^{(31)} = {1 \over 2\ell+1} \sum_m {\rm Real}\{ [\kappa^3]_{\ell m} [\kappa]^*_{\ell m} \}
= \sum_{\ell_1\ell_2\ell_3L} T^{\ell_3\ell}_{\ell_1\ell_2}(L)   J_{\ell_1\ell_2 L}J_{L\ell_3\ell}; \label{eq:defK31} \\
&& {\cal K}_{\ell}^{(22)} = {1 \over 2\ell+1} \sum_m  \{ [\kappa^2]_{\ell m} [\kappa^2]^*_{\ell m} \} 
= \sum_{\ell_1\ell_2\ell_3\ell_4} T^{\ell_3\ell_4}_{\ell_1\ell_2}(\ell)  J_{\ell_1\ell_2\ell}J_{\ell_3\ell_4\ell} \label{eq:defK22}.
\een
\ees
Thus the kurt-spectra described above are computed using either by keeping the diagonal
fixed and summing over all possible configurations (the two-to-two kurt-spectra ${\cal K}^{(2,2)}_{\ell}$
defined in Eq.(\ref{eq:defK31}))
or by keeping one of the side fixed and summing over all possible configurations
(introduced above as three-to-one kurt-spectra ${\cal K}^{(31)}_{\ell}$ defined in Eq.(\ref{eq:defK22})).
These {\em states} are linked to the collapsed and squeezed configurations. At higher-order the
polyspectra are characterized by a polygon. The number of polyspectra at a given order can be high
since the number of diagonals and sides of such polygons can be be quite high.

Another related point is that disconnected contributions will exists even in the absence of noise.
These contributions needs to subtracted out when estimating from the data \citep{Hu,OkamotoHu}.
The trispectrum in this case is given in Eq.(\ref{eq:Gauss}) and is specified completely by the
power spectrum ${\cal C}_{\ell}$. The corresponding spectra are given in terms of the Gaussian Trispectrum $G^{l_1l_2}_{l_3l_4}(L)$ \citep{xn1}:
\ben
  {\cal G}_{\ell}^{(31)} = \sum_{\ell_1\ell_2\ell_3L} G^{\ell_3\ell}_{\ell_1\ell_2}(L)   J_{\ell_1\ell_2 L}J_{L\ell_3\ell}; \label{eq:defG31} \quad
  {\cal G}_{\ell}^{(22)} = \sum_{\ell_1\ell_2\ell_3\ell_4} G^{\ell_3\ell_4}_{\ell_1\ell_2}(\ell)  J_{\ell_1\ell_2\ell}J_{\ell_3\ell_4\ell} \label{eq:defG22}.
\een
where the Gaussian trispectrum $G^{l_1l_2}_{l_3l_4}(L)$ is given by \citep{Hu,OkamotoHu}: 
\ben
&& G^{l_1l_2}_{l_3l_4}(L) = (-1)^{l_1+l_3} \sqrt {(2l_1+1)(2l_3+1)} {\cal C}_{l_1}{\cal C}_{l_3}\delta_{L0}\delta_{l_1l_2}\delta_{l_2l_3} \nn  \\
&& \hspace{2cm} + (2L+1){\cal C}_{l_1}{\cal C}_{l_2}\big [ (-1)^{l_2+l_3+L} \delta_{l_1l_3}\delta_{l_2l_4} + \delta_{l_1l_4}\delta_{l_2l_3}\big ].
\label{eq:Gauss}
\een
In \citep{MunshiHeavens,xn1,xn2} optimal versions of skew- and kurt-spectra estimators were developed which requires weights based on target spectra.
This method was used in investigating primordial spectra as the signal-to-noise is rather low. However,
for investigating the gravity induced secondary non-Gaussianity with surveys that have as high expected signal-to-noise as Euclid
optimization is not mandatory. 

The commonly used kurtosis parameter $S_4$ (to be defined below) can be reconstructed from the kurt-spectra as follows \citep{xn1}:
\bes
\ben
&& \mu_4 = \la \kappa^4(\oh) \ra_c = {1 \over 4\pi} \int \kappa^4(\oh) d\oh =
    {1\over 4\pi} \sum_L\sum_{\ell_1\ell_2\ell_3\ell_4} h_{\ell_1\ell_2L}h_{\ell_3\ell_4L}
T^{\ell_1\ell_2}_{\ell_3\ell_4}(L); \\
&& \hspace{2cm}  = \sum_{\ell} (2\ell+1){\cal K}^{(31)}_{\ell} =  \sum_{\ell} (2\ell+1){\cal K}^{(2,2)}_{\ell}.
\een
\ees
We will use noise free simulations but in case of analyzing noisy maps the ${\cal C}_{\ell}$s will also include the noise contribution.
The commonly used kurtosis are normalized one-point estimators as \citep{bernardeau_review}  $ S_4 = \left [ {\mu_4 - 3\mu_2^2 \over \mu_2^3} \right ]$.
Here, $\mu_2 = {1/4\pi} \sum_\ell (2\ell+1) {\cal C}_{\ell}$.
The corresponding cumulant correlators for these spectra are defined in a manner similar to Eq.(\ref{eq:corr21}) \cite{xn2}:
\bes
\ben
&& \xi^{31}(\theta_{12}) = \langle\kappa^3(\oh_1)\kappa(\oh_2)\rangle_c =
   {1 \over 4\pi} \sum_{\ell} (2\ell+1) {\cal K}^{(31)}_{\ell}P_{\ell}(\cos\theta_{12});  
   \label{eq:corr31} \\
&&   \xi^{22}(\theta_{12}) = \langle\kappa^2(\oh_1)\kappa^2(\oh_2)\rangle_c =
   {1 \over 4\pi} \sum_{\ell} (2\ell+1) {\cal K}^{(22)}_{\ell}P_{\ell}(\cos\theta_{12}).  
   \label{eq:corr31}
   \een
   \ees
Next we will employ tree-level perturbative calculations.
%
\section{Tree-level Perturbative Results}
\label{sec:tophat}
The unsmoothed normalized higher-order cumulants or $S_N = \langle \delta^N \rangle_c/\langle \delta^2 \rangle^{N-1}_c$
can be expressed in terms of the tree-level vertices denoted as $\nu_N$ using the following expressions \citep{bernardeau_review}:
\ben
&& S_3 = 3\nu_2; \quad
S_4 = 4\nu_3+ 12\nu_2^2; \quad
S_5 = 5\nu_4+ 60\nu_3\nu_2 + 60\nu_2^3. \quad
\label{eq:SN}
\een
The vertices $\nu_N$ are the angular averages of the mode-coupling kernels $F_N$  defined in Eq.(\ref{eq:F3}) i.e. $\nu_N=N!\la F_N \ra$
introduced in \textsection\ref{sec:flat_sky} in the Fourier domain.
\ben
&& \nu_N = N!\langle F_N \rangle=  N! \int {d\oh_{k_1} \over 4\pi} \cdots \int {d\oh_{k_N} \over 4\pi} F_N(\bk_1.\cdots, \bk_N);\quad d\oh_k = \sin\theta_k d\theta_k d\varphi_k.
\label{eq:define_nuN}
\een
The following generating function approach was introduced in \citep{Bernard92, Bernard1}.
  The generating functions ${\cal G}_{\delta}(\tau)$ are solved using the equations of gravitational dynamics encapsulated in
  Euler-Continuity-Poisson equations. Here $\tau$ plays the role of a dummy variable.
  In the perturbative regime the $\nu_N$ parameter can be computed for an arbitrary $N$.  
  \ben
  && {\cal G}_{\delta}(\tau) = \sum_{n}{\nu_N \over N!}\tau^N
    = -\tau + {12 \over 14} \tau^2 - {29 \over 42} \tau^3 + {79 \over 147} \tau^4 - {2085 \over 5096} \tau^5 + \cdots
    \een
    Next, using Eq.(\ref{eq:SN}), the one-point cumulants in 2D \citep{MBMS99}, denoted as $\Sigma_N$ as opposed to $S_N$ parameters which
    represent the cumulants in 3D, can be used to compute the cumulants to arbitrary order in 2D \citep{MBMS99}.
  \ben
  && \Sigma_3 = {36\over 7}; \quad  \Sigma_4={2540 \over 49}; \quad  \Sigma_5={793}; \quad  \Sigma_6=16370;
  \een  
  The generalization of the one-point cumulants i.e the $S_N$ parameters to the two-point cumulant correlators
  $C_{pq} = \la\delta^p_1\delta_2^q\ra_c/\la\delta^2\ra_c^{p+q-1}\la\delta_1\delta_2\ra_c$
  or $C_{pq}$ parameters was introduced
  in \citep{Bernardeau_bias}. The lower-order normalized cumulant correlators can also be expressed
  in terms of the tree-level vertices $\nu_N$ just as
  the one-point cumulants introduced in Eq.(\ref{eq:SN}).
\ben
&& C_{21} = 2\nu_2; \quad C_{31} = 3\nu_3+ 6\nu_2^2; \quad C_{41} = 4\nu_4 + 36 \nu_3\nu_2  + 24 \nu_2^3.   
\een
To compare with observed or simulated data smoothing of the field is necessary. 
The smoothed generating function ${\cal G}^s_{\delta}$ can be computed from the unsmoothed generating function ${\cal G}_{\delta}$.
The generating functions ${\cal G}^s_{\delta}$  and ${\cal G}_{\delta}$ are related by the following implicit relation \citep{B95}
\ben 
&& {\cal G}_{\delta}^s(\tau)= {\cal G}_{\delta}(\tau[1+ {\cal G}_{\delta}^s]^{-(2+n)/4}).
\label{eq:implicit2D}
\een
A {\em tophat} smoothing window is assumed and the power spectrum is approximated locally as a power law $P(k)\propto k^n$ \citep{MBMS99,B95}.
For other window functions, e.g. Gaussian window generic results are not possible for arbitrary $N$. However, an order-by-order
approach can be adopted to obtain the lower-order cumulants \citep{Matsubara02}.
Notice that the smoothed power law  depends on the spectral index while unsmoothed vertices depend solely on the gravitational collapse
in 3D spherical or cylindrical in 2D. The smoothed vertices can be recovered by Taylor-expanding the smoothed generating function ${\cal G}^s$.
Using these vertices it is possible to now compute the 2D skewness $\Sigma_3$ and kurtosis $\Sigma_4$ can be computed \citep{MBMS99}:
\bes
\ben
&& \Sigma_3 = {36 \over 7} -{3\over 2} (n+2); \label{eq:2DS3}\\
&& \Sigma_4 = {2540 \over 49} - 33(n+2) + {21 \over 4}(n+2)^2. \label{eq:2DS4}
\een
\ees
These expressions are derived using 2D where gravitational collapse with cylindrical symmetry is relevant as is the case for projected
surveys. However, the underlying statistics for the 3D density field is linked to spherical collapse which we have not considered here but
may be relevant for a 3D weak lensing scenario where photometric data is used. However, there is a crucial difference between 2D and 3D
statistics.
For large separations in 3D we can factorise $C_{pq}=C_{p1}C_{q1}$, while in 2D this approximation is not valid.
%
%
Thus, we will consider the family of statistics $C_{p1}$ for arbitrary $p$.
\bes
\ben
&& {\cal S}^{(21)}_{\ell} = R_2 \Sigma_{21} P^{\kappa}(l)\sigma_L^2  = R_2 \left [ {24 \over 7} - {1 \over 2}(n+2)
  \right ] P_{\delta}(l)\sigma_L^2; 
\quad \sigma_L^2=\la\kappa^2\ra; \label{eq:define_B} \\
&& R_2 = \int_0^{r_s} d\,r {w^3(r)\over d_A^{4+2n}(r)} \bigg / \left ( \int_0^{r_s} d\,r{ w^2(r) \over d_A^{2+n}(r)} \right )^2.
\label{eq:ib_def}
\een
\ees
The corresponding result at the fourth-order is given by:
\bes
\ben
&& {\cal K}^{(31)}_{\ell} = R_3 \Sigma_{31} P^{\kappa}(l)\sigma_L^4 = R_3 \left [ {1473\over 49} - {195\over 14}(n+2) + {3 \over 2}(n+2)^2 \right ]
 P_{\delta}(l)\sigma_L^4 ; \label{eq:defineK}\\
 && R_3 = \int_0^{r_s} d\,r {w^4(r)\over d_A^{6+3n}(r)} \bigg / \left ( \int_0^{r_s} d\,r{ w^2(r) \over d_A^{2+n}(r)} \right )^3.
 \label{eq:iK_def}
 \een
 \ees
 The dynamical contribution is encoded on $\Sigma_{p1}$ where as the line-of-sight integration is represented by the
 pre-factors in $R_{p}$.
 
 Historically the generating function
 approach was developed without any reference to perturbative dynamics and the vertices were left undetermined.
 Many generic predictions were developed coupling scaling Ans\"atze with the generating function formalism \citep{BS89}.
 While in the quasi-linear regime the loop corrections to the tree-level results violate the scaling Ansatz,
 in the highly nonlinear regime the vertices are known to become shape independent parameter
 as encapsulated in Hyper Extended Peturbation Theory \citep{Scoccimarro}.
 In recent years some of the results were derived the Large Deviation Principle 
 \citep{BernardeauReimberg16, Uhlemann15, ReimbergBernardeau, Uhlemann18}.

 Previously, many studies have focused on observed and simulated data of one-point cumulants \citep{B95, GB98}
 as well as for the two-point cumulant correlators \citep{MCM,SS99}. Previous studies have focused on galaxy surveys.
 In this paper we extend these results to the context of weak lensing.
 %
\begin{figure}
  \centering
  \includegraphics[width=8cm]{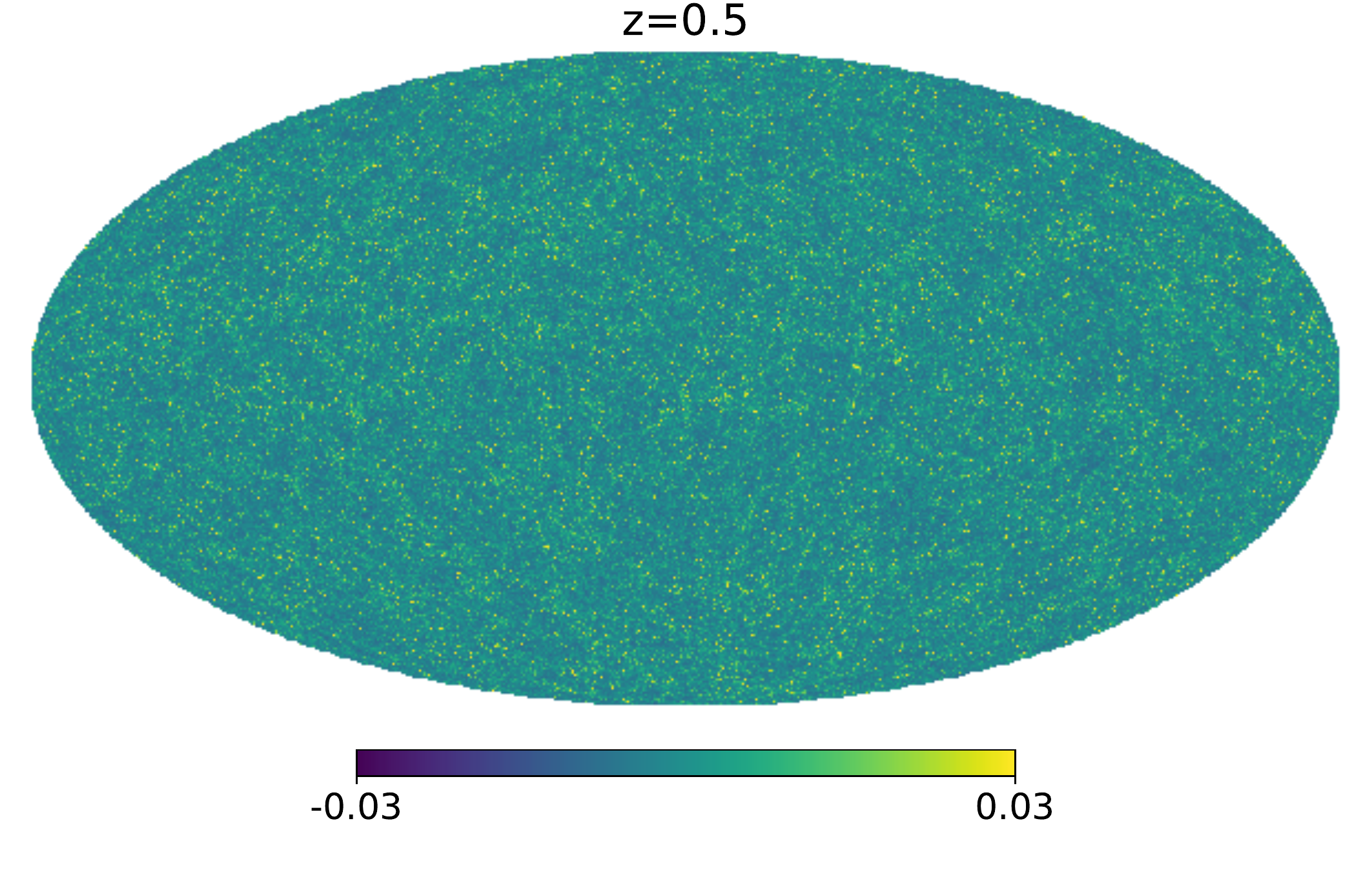}
  \hspace{1cm}
   \includegraphics[width=8cm]{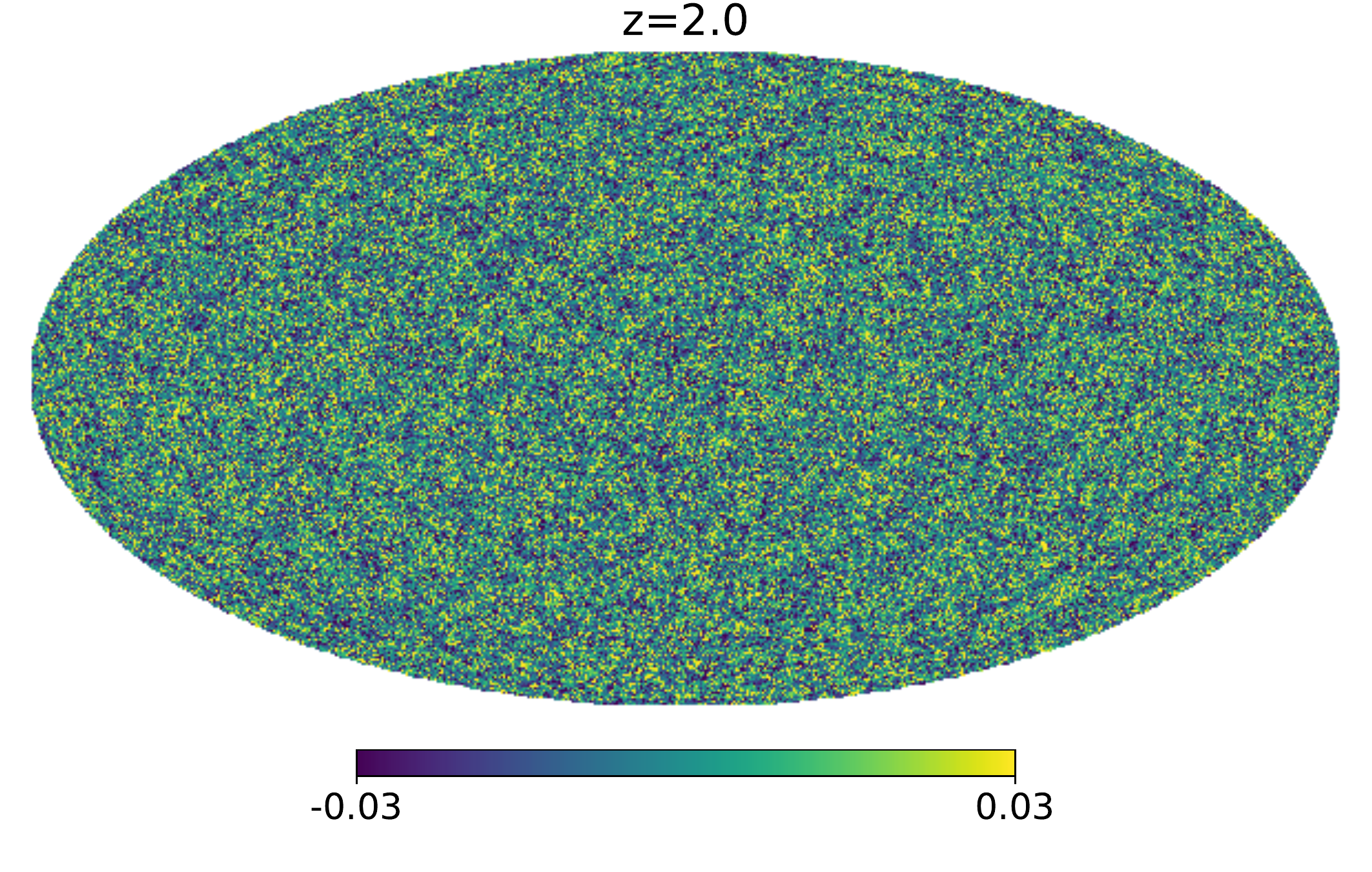}
  \caption{Examples of simulated $\kappa$ maps used in our study. The left panel corresponds to $z_s = 0.5$ while the right panel corresponds to
    $z_s = 2.0$. The maps were generated at a resolution of $N_{\rm side} = 4096$. See \textsection\ref{sec:simu} for more detail
    discussion about construction of maps used in our study.}
  \label{fig:maps}
\end{figure} 
%
\section{Higher-Order Spectra In Three Dimensions}
\label{sec:3D}
Next, we will consider higher-order statistics in three dimensions (3D).
Future surveys such as Euclid will go beyond the projection and
using photometric redshifts will be able to retain radial information.
In 3D we will compute the higher-order spectra as before in the low-$\ell$
limit. The results have similar characteristics as in projection, which
we have discussed in \textsection\ref{sec:tophat} but are very different in
certain aspect as we discuss below. We will decompose the lensing field in
two different eigenmodes (a) Fourier-Bessel decomposition typically used
for radially symmetric fields and (b) The generic Fourier-Cartesian decomposition
that are most commonly used for perturbative analysis.

We will use follow the same convention for forward and reverse Fourier transformation introduced in Eq.(\ref{eq:Fourier_cartesian}) for the
Cartesian co-ordinate. For an arbitrary function $\rA(\br)$ with $\br\equiv(r,\oh)=(r,\theta,\phi)$ and its Fourier tarnsform  $\rA({\bf k}; r)$ we will use:
\ben
&& \rA({\bf r}; r) =  \int d{\bk}\, \rA(\bk; r)\, \exp(i\bk\cdot\br); \quad
\rA({\bf k}; r) = \int {d\br}\, \rA(\br; r)\, \exp(i\bk\cdot\br).
\een
In spherical-Bessel coordinates the eigenfunctions of the Laplacian operators are the products of spherical harmonics $Y_{\ell m}(\oh)$ and
spherical Bessels function $j_{\ell}(r)$ i.e. $j_{\ell}(kr)Y_{\ell m}(\oh)$ the transforms take the following form:
\ben
&& \rA_{\ell m}(k) \equiv  \sqrt{2\over \pi}\int d^3{\br} \rA(\br) k j_{\ell}(kr)Y_{\ell m}^{*}(\oh); \quad
\rA(\br) \equiv \sqrt{2\over \pi} \int k dk \sum_{\ell =0}^{\infty}\sum_{m-\ell}^{\ell} \rA_{\ell m}(k)j_{\ell}(kr)Y_{\ell m}(\oh).
\een
Using the well-known Rayleigh expansion that expands the plane-wave in a spherical wave basis: 
\ben
&& \exp(i\,\bk\cdot \br) = 4\pi \sum_{\ell}\sum_{m=-\ell}^{m=\ell} i^\ell j_{\ell}(kr) Y_{\ell m}(\oh_k)Y_{\ell m}(\oh);  \quad\quad \oh_k=(\theta_k,\phi_k).
\een
we can relate the spherical harmonic coefficients $\rA_{\ell m}$ with their Fourier counterpart $\rA$:
\ben
&& \rA_{\ell m}(k;r) = {1 \over ({2\pi})^{3/2}}\, k\, i^\ell\, \int\, d\oh_k\, \rA(\bk; r) Y_{\ell m}(\oh_k).
\een
§§The 3D power spectrum $P^{\rm AA}(k)$ defined respectively in Cartesian coordinates and
${\cal C}_{\ell}^{\rm AA}$ in spherical coordinates are:
\ben
&& \langle\rA(\bk)\rA^*(\bk)\rangle = (2\pi)^3P^{AA}(k); \quad
\langle \rA_{\ell m}(k,l)\rA_{\ell^{\prime} m^{\prime}}^*(k^{\prime}) \rangle = (2\pi)^2 {\cal C}^{\rm AA}_{\ell}(k; r) \delta_{\rm 1D}(k-k^{\prime})\delta_{\ell\ell^{\prime}}\delta_{mm^{\prime}}. \quad 
\een
In general, in the absence of any mask, it can be shown that: ${\cal C}_{\ell} = P(k)$ i.e. the
3D power spectrum in spherical co-ordinate is independent of $\ell$ and is actually same
as the power spectrum in Cartestian co-ordinates \citep{Castro05}. 
Next, for the construction of the higher-order 3D spectra we will define the
following cross-spectra between two arbitrary 3D fields $\rA(\br)$ and $\rB(\br)$
in spherical co-ordinates: 
\ben
&& \langle\rA(\bk)\rB^*(\bk)\rangle = (2\pi)^3P^{AB}(k); \quad
\langle \rA_{\ell m}(k; r) \rB^*_{\ell m}(k^{\prime};r) \rangle = (2\pi)^2 {\cal C}^{\rm AB}_{\ell}(k; r) \delta_{\rm 1D}(k-k^{\prime})\delta_{\ell\ell^{\prime}}\delta_{mm^{\prime}}.
\een
Using this identity, for the 3D density field $\delta$ we can derive the following
expressions for the higher-order spectra of the density field:
\bes
\ben
\label{eq:c11}
&& P^\delta(k; r,\rp) = \langle \delta(\bk;r)\delta^{*}(\bk',\rp) \rangle_c;
\quad\quad {\cal C}^{\delta}_{\ell}(k; r,\rp) = \langle \delta^{}_{\ell m}(k;r)\delta^{*}_{\ell m}(k;\rp)\rangle_c;\quad
P^\delta(k; r,\rp) = {\cal C}^{\delta}_{\ell}(k; r,\rp). \label{eq:cumu1} \\
&& S^{21,\delta}(k;r,\rp) = \langle\delta^{2}(\bk;r)\delta^{*}(\bkp;\rp)\rangle_c \, ;
\quad\quad S^{21,\delta}_{\ell}(k; r,\rp) = \langle \delta^{2}_{\ell m}(k; r)\delta^{*}_{\ell m}(k;\rp)\rangle_c;\quad
 S^{21,\delta}(k,r,\rp) = S^{21,\delta}_{\ell}(k; r, \rp).\\
&& T^{31,\delta}(k;r,\rp) = \langle\delta^{3}(\bk;r)\delta^{*}(\bkp; \rp)\rangle_c \, ;
\quad T^{31,\delta}_{\ell}(k,r) = \langle \delta^{2}_{\ell m}(k;r)\delta^{*}_{\ell m}(k;\rp)\rangle_c ; \quad
 T^{31,\delta}(k,r,\rp) = T^{31,\delta}_{\ell}(k,r,\rp). \\
&& T^{22,\delta}(k,r,\rp) = \langle\delta^{2}(\bk;r)\delta^{2*}(\bk';\rp)\rangle_c \, ;
\quad T^{22,\delta}_{\ell}(k;r,\rp) = \langle \delta^{2}_{\ell m}(k;r)\delta^{2*}_{\ell m}(k;\rp)\rangle_c; \quad
T^{22,\delta}(k,r,\rp) = T^{22,\delta}_{\ell}(k; r,\rp).
\label{eq:c31}
\een
\ees
In our notation, $\delta^p(k)$ is the Fourier transform of $\delta^p$.
Notice these expressions are non-perturbative and are valid irrespective of detailed
modelling and are valid to an arbitrary order i.e. when cross-correlating p-th
power of $\delta$ against the q-th power i,e in $\langle\delta^{p}(\bk)\delta^{q}(\bk)\rangle$
in spherical or Cartesian co-ordinate.
In the Cartesian co-ordinate the normalized cumulant correlators $C_{pq}$ are defined as follows:
\ben
&& \langle\delta^{p}(\bk)\delta^{q*}(\bk)\rangle_c = C_{pq} \langle \delta^2 \rangle_c^{p+q-2} P(k) 
= C_{pq} \langle \delta^2 \rangle_c^{p+q-2} {\cal C}_{\ell}(k)
\label{eq:define_cpq_kspace}
\een
The second step relies on Eq.(\ref{eq:cumu1}). In the real-space Eq.(\ref{eq:define_cpq_kspace})
this is equivalent to:
\ben
&& \langle\delta^{p}(\br_1)\delta^{q*}(\br_2)\rangle_c =
C_{pq} \langle \delta^2 \rangle_c^{p+q-2} \langle\delta(\br_1)\delta(\br_2)\rangle_c.
\een
The results Eq.(\ref{eq:c11}) - Eq.(\ref{eq:c31}) are non-perturbative and do not depend on any simplifying
assumptions. However, in case of studies of galaxy clustering it is more natural to study high-order statistics in the redshift space.
Similarly, for 3D weak lensing, line-of-sight integration will need to be taken into account. 
Such extensions will be presented separately.
The coefficients $C_{pq}$ defined in Eq.(\ref{eq:define_cpq_kspace}) can be computed using perturbative calculations.
In 3D the smoothed and unsmoothed vertex generating functions are related through an implicit expression \citep{Bernardeau_bias}
which is analogous to Eq.(\ref{eq:implicit2D}). 
\ben 
&& {\cal G}_{\delta}^s(\tau)= {3 \over 2}{\cal G}_{\delta}(\tau[1+ {\cal G}_{\delta}^s]^{-(3+n)/6}).
\een
The power spectrum is assumed to be approximated locally by a power law with power law index $n$ i.e. $P(k)\propto k^n$.
On Taylor expanding the 3D (unsmoothed) generating function ${\cal G}(\tau)$, we can recover the lower order vertices $\nu_N$ in 3D \citep{bernardeau_review}:
\ben
  && {\cal G}_{\delta}(\tau) = \sum_{n}{\nu_N \over N!}\tau^N
= -\tau + {34 \over 21} \tau^2 - {682 \over 189} \tau^3 + \cdots
\label{eq:3Dnu}
    \een
Using these vertices it is possible to compute the normalized lower-order moments i.e. skewness $S_3$ and kurtosis $S_4$ in 3D \citep{bernardeau_review}:
\bes
\ben
&& S_3 = {34\over 7} -{}(n+3); \quad\quad S_4 = {60712 \over 1323} - {62 \over 3}(n+3) + {7\over 3}(n+3)^3.
\label{eq:3DS3}
\een
The lower-order cumulant correlators have the following form \citep{bernardeau_review}::
\ben
&& C_{21} = {68\over 21} -{(n+3)\over 3}; \quad\quad
C_{31} = {11710 \over 441} - {61 \over 7}(n+3) + {2\over 3}(n+3)^3.
\label{eq:3DC21}
\een
\ees
Detailed derivations regarding construction of one- and two-point proabability distribution
functions are detailed in \citep{bernardeau_review}.
The 3D vertices defined in Eq.(\ref{eq:3Dnu}) assume a different numerical value though the formal
structure remains the same. In addition a more generic results $C_{pq}=C_{p1}C_{q1}$
gives a much needed consistency check.
The results in a 3D collapse are related to a spherical window and the dynamics
relate to the 3D spherical collapse. 
%
%
%
\section{Pseudo-${\cal C}_{\ell}$ (PCL)  Estimators}
\label{sec:pcl}
%
%
\begin{figure}
  \centering
  \includegraphics[width=5cm]{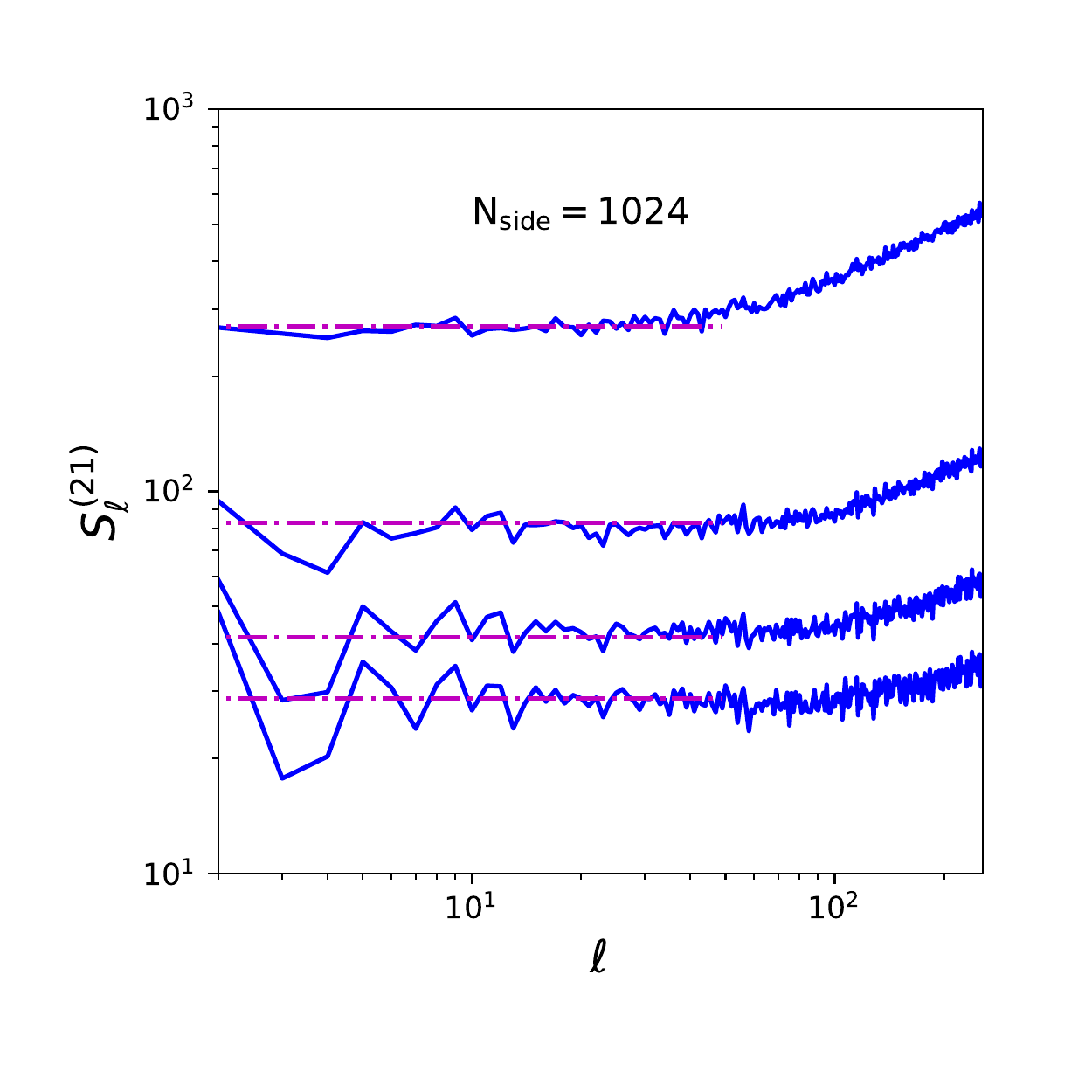}
   \caption{The skew-spectrum $S^{(21)}_{\ell}$ defined in Eq.(\ref{eq:defSkew})-Eq.(\ref{eq:defJ}) is shown as a function of the harmonics $\ell$. 
     From top to bottom the curves correspond to source redshifts $z_s=0.5,1.0,1.5$ and $2.0$ respectively.
     A total of 10 simulations were used to
   compute the $S^{(21)}_{\ell}$. The straight lines at the left correspond to predictions from perturbation theory encapsulated in
   Eq.(\ref{eq:define_B})- Eq.(\ref{eq:ib_def}.
   We have assumed a power-law power spectrum $P_{\delta}(k)\propto k^n$.
   We have chosen $n=-2.0$ (dot-dashed lines).  See text for details.}
   %
  \label{fig:s21}
\end{figure} 
%
Maximum likelihood (ML) estimators or quadratic maximum likelihood (QML) estimators are
most obvious choices for analyzing cosmological data sets. However, these estimators
require inverse covariance weighting which clearly is not practical
for large cosmological data sets though various clever algorithmic techniques have been considered \citep{bernardeau_review}.
This has resulted in the development of
many sub-optimal estimators which uses heuristic  weighting schemes.
The so-called pseudo-$\cal C_{\ell}$ (PCL) technique was introduced in \citep{Hivon_Master}; 
see \cite{Spice} for a related method. These estimators are unbiased but sub-optimal.
Various weighting schemes depending on sky coverage as well as noise characteristic as
well as various hybridization schemes to combine large angular scale (equivalently the low $\ell$) estimates
using QML with small angular scale (high $\ell$) PCL estimates.
were considered in \citep{GPE}.
\bes
\ben
&& M_{\ell\ell'} = (2\ell'+1)\sum_{\ell''}
\left ( \begin{array}{ c c c }
     \ell & \ell' & \ell'' \\
     0 & 0 & 0
  \end{array} \right)^2
{ (2\ell'' +1 )\over 4\pi} |w_{\ell''}^2|; \label{eq:MLL1}\\
&& \hat{\cal S}_{\ell}^{(21)} = \sum_{\ell'}M_{\ell\ell'}^{-1} \tilde{\cal S}_{\ell'}^{(21)} \label{eq:MLL2}.
\een
\ees
Here $\tilde{\cal S}_{\ell'}^{(21)}$ denotes the skew-spectrum computed from a map in the presence of a mask $w(\oh)$,
$\hat{\cal S}_{\ell'}^{(21)}$ is the all-sky estimate and $w_{\ell} = {1/(2\ell+1)} \sum_m w_{\ell m} w^*_{\ell m}$ is the
power spectrum of the mask constructed from the harmonic-coefficient $w_{\ell m}$ of the map. The {\em coupling} matrix $M_{\ell\ell^{\prime}}$
is represents the mode-mixing due to the presence of a mask.
The generalization of the PCL method to estimate higher-order spectra were developed in \citep{xn1,xn2} 
for spin-0 fields and in higher spin fields in \citep{flexions} as well as in 3D in \citep{MunshiKitching}.
Exactly same result holds for higher-order spectra, e.g., for all-sky estimate of kurt-spectrum $\hat{\cal K}_{\ell}^{(21)}$ and
its masked counterpart $\tilde{\cal K}_{\ell'}^{(21)}$ are related through a similar expression 
$\hat{\cal K}_{\ell}^{(31)} = \sum_{\ell'}M_{\ell\ell'}^{-1} \tilde{\cal K}_{\ell'}^{(31)}$.
This has also been generalized to reconstruct the Minkowski Functionals in an order-by-order manner \citep{waerbeke,cmb_minkowski}
Two equivalent techniques for flat-sky
PCLs are developed here \citep{Marika} and \citep{Chiaki1}.
%
\section{Numerical Simulations}
\label{sec:simu}
%
%
We use the publicly available all-sky weak-lensing maps generated by \citep{Ryuchi}\footnote{http://cosmo.phys.hirosaki-u.ac.jp/takahasi/allsky\_raytracing/}
that were generated using ray-tracing through N-body simulations.  
Multiple lens planes were used to generate convergence ($\kappa$) as well as shear ($\gamma$) maps.
Many recent studies were performed using these maps, e.g. \cite{Namikawa18,MunshiNamikawa19}.
In these simulations, the source redshifts used were in the range $z_s= 0.05-5.30$ at interval $\Delta z_s = 0.05$.
In this study, we have used the maps with $z_s=0.5,1.0,1.5, 2.0$.
The maps do include post-Born corrections \citep{LewisPratten}.
Though at the low source redshift such corrections only play a negligible role. Indeed, they do play significant role
in CMB lensing. The convergence maps were generated using an equal area pixelisation scheme.
in {\tt HEALPix}\footnote{https://healpix.jpl.nasa.gov/} format\citep{Gorski}.
In this pixelisation scheme the number of pixels scale as $N_{\rm pix} = 12 N^2_{\rm side}$
where $N_{\rm side}$ is the resolution parameter which can take values $N_{\rm side} = 2^{N}$ with $N=1,2,\cdots$.
The set of maps we use are generated at $N_{\rm side}=4096$ which were also cross-checked using higher resolution maps
that were constructed at a resolution $N_{\rm side}=8192, 16384$. These maps were found to be
consistent with each other up to the angular harmonics $\ell \le 3600$. In addition detailed
tests were performed by using a Electric/Magnetic (E/B)  decomposition of shear maps
for the construction of $\kappa$ maps \citep{Ryuchi}.
Though we have used high resolution maps $N_{\rm side} =4096$,  we have degraded them to
low resolution maps at $N_{\rm side}=1024$ as we are primarily interested in the perturbative regime. 
The following set of cosmological parameters $\Omega_{\rm CDM} = 0.233$, $\Omega_b = 0.046$,
$\Omega_{\rm M} = \Omega_{\rm CDM}+\Omega_b, \Omega_{\Lambda}=1-\Omega_{\rm M}$ and $h=0.7$ were used to generate the maps
assuming a $\Lambda$CDM background cosmology
The amplitude of density fluctuations $\sigma_8=0.82$ and the spectral index $n_s=0.97$.
Examples of $\kappa$ maps used in our study are presented in Figure-\ref{fig:maps}
We will be focus on the large-separation or the small $\ell$ regime in our study
and we do not expect the baryonic feedback to play a significant role \citep{Weiss19}.
It is worth mentioning here that these maps were also used to recently analyze the bispectrum
the context of CMB lensing \citep{Namikawa18}. 
%
%
\section{Tests Against Numerical Simulations}
\label{sec:num}
%
\begin{figure}
  \centering
  \includegraphics[width=5cm]{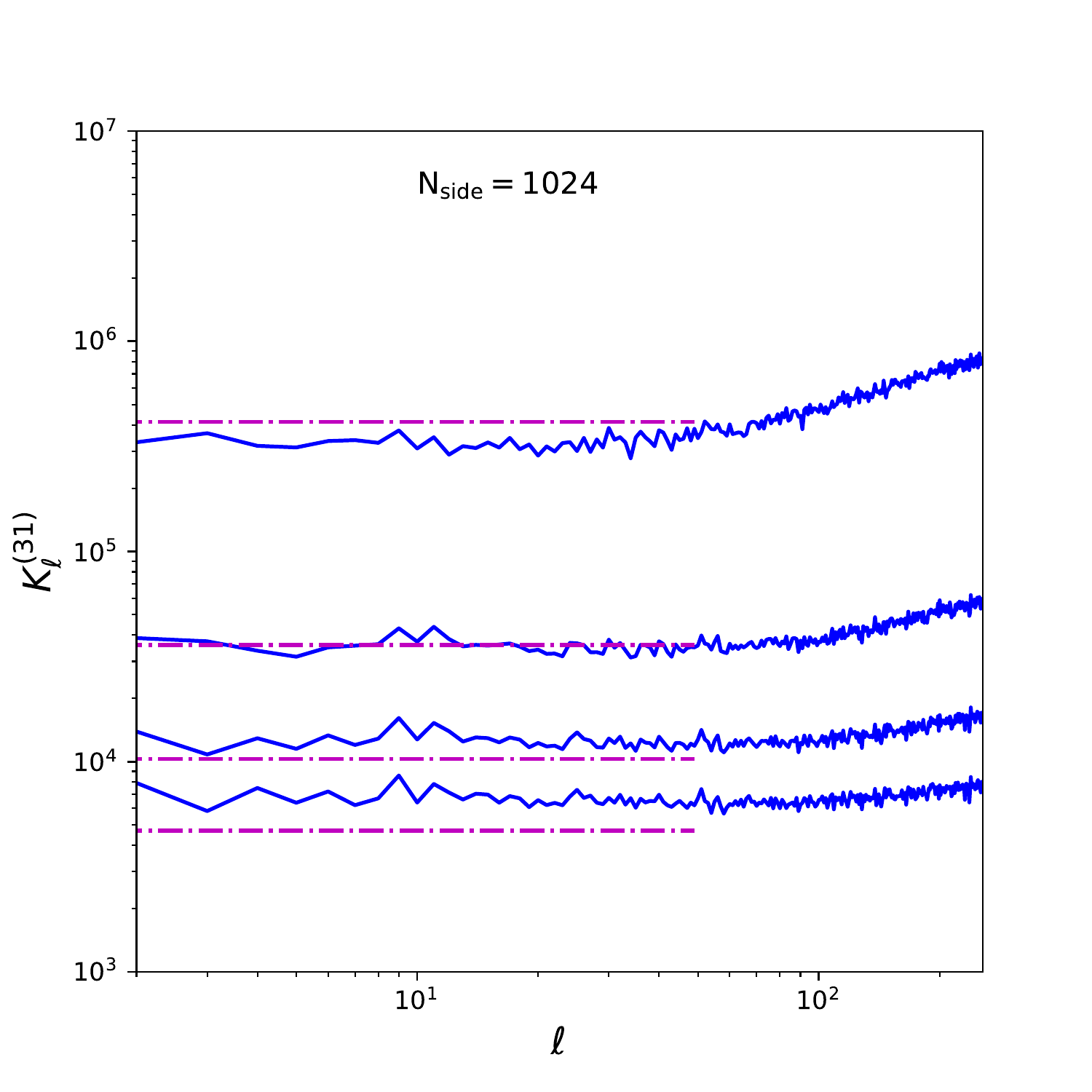}
  \caption{The skew-spectrum $K^{(31)}_{\ell}$ defined in Eq.(\ref{eq:defK31}) is shown as a function of the harmonics $\ell$. 
    From top to bottom the curves correspond to source redshifts $z_s=0.5,1.0,1.5$ and $2.0$ respectively.
    A total of 10 simulations were used to
   compute the $K^{(31)}_{\ell}$. The straight lines at the left correspond to predictions from perturbation theory encapsulated in
   Eq.(\ref{eq:defineK})-Eq.(\ref{eq:iK_def}).
   We have assumed a power-law power spectrum $P_{\delta}(k)\propto k^n$.
   We have chosen $n=-2$. See text for details.}
  \label{fig:k31}
\end{figure} 
%
The skew-spectrum ${\cal S}^{(21)}_{\ell}$ is shown as a function of the harmonics $\ell$ in Figure-\ref{fig:s21}. From top to bottom the curves represents
the source redshifts $z_s = 0.5, 1.0, 1.5$ and
$2.0$ respectively. The results are from maps with $N_{side} = 1024$. 
We have analyzed these maps for $\ell_{max} = 2N_{side}$. The
straight lines correspond to perturbative results computed using tree-level perturbation theory 
Eq.(\ref{eq:define_B}) - Eq.(\ref{eq:ib_def}). We have used an ensemble of ten realisations to compute the mean which is being plotted. We use all-sky maps without
an observational mask. The effect of mask can be incorporated using Eq.(\ref{eq:MLL1})-Eq.(\ref{eq:MLL2}).

The ${\cal K}^{(31)}_{\ell}$ is shown as a function of the harmonics $\ell$ in Figure-\ref{fig:k31}.
From top to bottom the curves represents the source redshifts $z_s = 0.5, 1.0, 1.5$ and
$2.0$ respectively. The maps used are $N_{side} = 1024$ and as before we have analyzed for $\ell_{max} = 2 N_{side}$.
The straight lines corresponds to perturbative results computed using tree-level perturbation theory  Eq.(\ref{eq:defineK}) - Eq.(\ref{eq:iK_def}). We have used an ensemble
of ten realisations to compute the mean which is being plotted. We use all-sky maps without an observational mask.

Our results for the skew- and kurt-spectra are derived in the large separation limit i.e. the cumulant correlators
defined, e.g. in Eq.(\ref{eq:corr21}) and in Eq.(\ref{eq:corr31}) $|\xi_{12}|/\bar\xi_2 \ll 1$.
In real-space this limit was seen to be reached very fast as soon as the two neighboring cells are not overlapping.
In harmonic domain the scale $\ell$ represents the separation of two beam-smoothed pixels for which
the skew-spectrum is being measured. Thus, large separation in our case corresponds to low $\ell$, and
typical size of the pixels corresponds to the $\ell$ at which the beam can no longer be approximated as unity.
This is the scale where the correction to the skew-spectrum starts to be non-negligible. These corrections,
which are of order $\xi_{ij}/\bar\xi_2 \ll 1$, are difficult to compute analytically. Though, entire
skew-spectrum can be computed with fitting functions. Clearly, such a computation will not be possible
beyond third-order i.e. skew-spectrum due to lack of such a fitting function at the fourth-order.
Thus, the techniques developed here are valuable as their
predictions can be made at {\em all-orders}.  

The results we have computed are based on spherical top-hat window. However, many previous studies
have shown that the actual shape of the window is not important, and replacing the circle with square
can be very good approximation \citep{MBMS99}. However, profile of the smoothing beam or window
as opposed to its shape can change the theoretical predictions. The predictions for a Gaussian
window was worked out in detail in \citep{Matsubara02}. However, results can be derived only in order-by-order
manner and approaches based on generating functions are not applicable.

A few comments about going beyond kurt-spectrum are in order.
Extraction and interpretation of higher-order statistics can be rather complex from any cosmological data-sets.
Estimators of the cumulants and cumulant correlators are typically known to be biased and elaborate scheme
were developed in estimating and correcting such bias as well as scatter in estimators typically used for evaluating these quantities
mainly in real space in the context of galaxy clustering. Such corrections are expected to be more dominant role
with increasingly higher-order \citep{MBMS99}. Such corrections and their accurate calibration against simulations are
lacking in the literature. Though, for lower-order statistics we probed here,
such corrections are expected to be negligible, a better understanding of such effects is needed before
we can interpret the statistics beyond kurt-spectra (equivalently trispectrum).

An alternative approach considered by various authors was to consider the one-point and two-point
probability functions which encode cumulants an their correlators to an arbitrary order \citep{MunshiJain2,Munshi_bias}.
These results are applicable in real-space which make them useful for surveys with low-sky coverage.
The results derived here will be relevant for surveys with high sky-coverage where
harmonics decomposition would mean less correlated measurements for individual $\ell$.

By their very nature, projected or 2D surveys unlike their 3D counterparts mixes scale which makes assigning exact spectral index
with an angular scales or in our case the harmonic $\ell$. We have shown how much variation we should expect for a range of feasible spectral index $n$.
Finally, the redshift dependence of the skew- and kurt-spectra is encoded in the coefficients $R_2$ and $R_3$. It is
however important to point out that these pre-factors are rather sensitive to the lower limit of the integration $z_{\rm min}$ i.e. in  Eq.(\ref{eq:ib_def}) and Eq.(\ref{eq:ib_def}).
Numerical implementation of simulation of ray-tracing to generate convergence maps may introduce slight modification in $z_{\rm min}$ which may lead to
a bias in the theoretical predictions. 

The excellent match between the theoretical predictions and simulations we have found here is encouraging for computing such corrections.
%
\section{Modified Theories of Gravity: Computation of $C_{21}$}
\label{sec:modG}
%
\begin{figure}
  \centering
  \includegraphics[width=6cm]{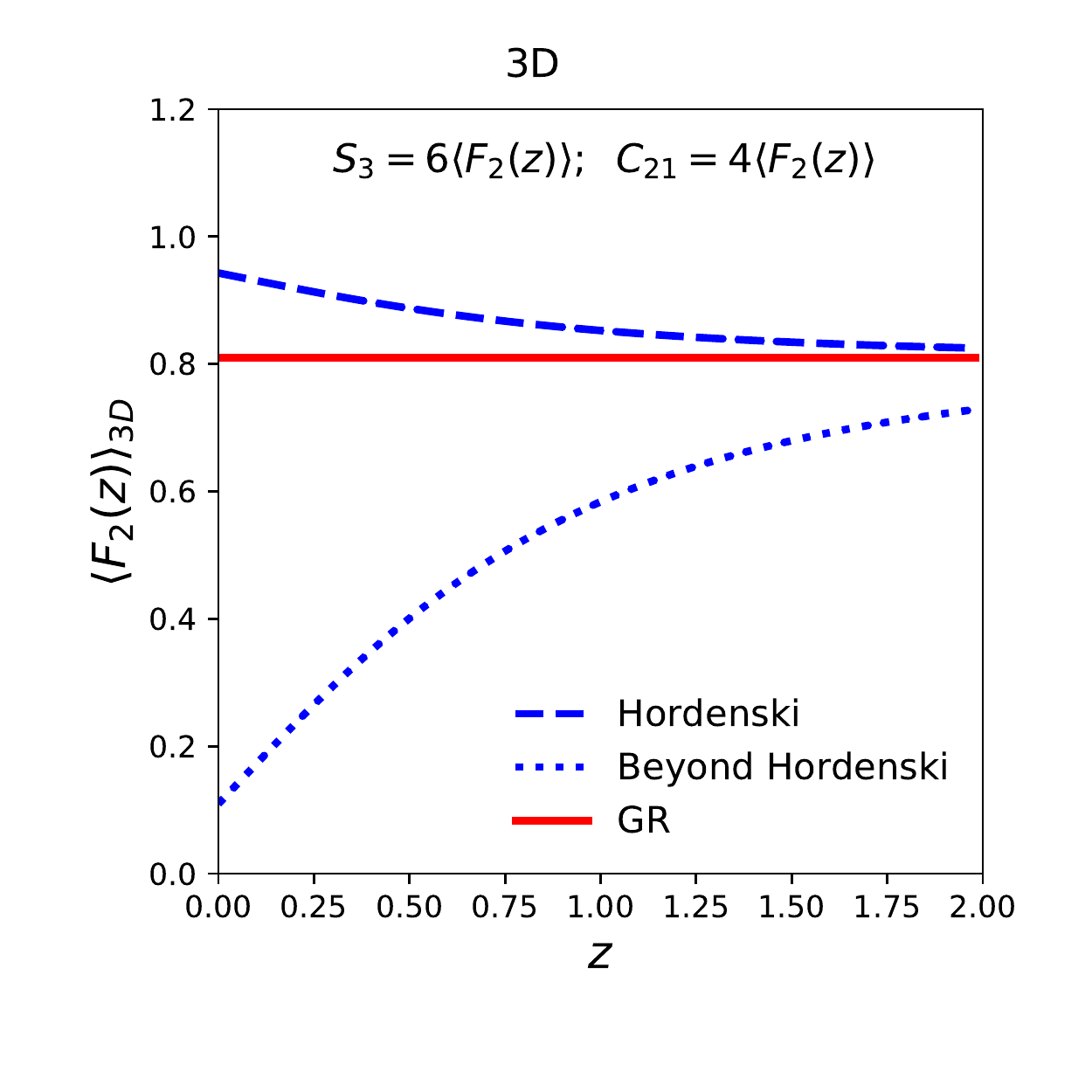}
  \caption{The second-order tree-level perturbative vertex $\langle F_2(z) \rangle_{\rm 3D}$ is plotted as a function of redshift $z$
    for 3D surveys as given in Eq.(\ref{eq:2Dand3D}).
    Three different cases shown correspond to the Horndeski, Beyond Horndeski and GR as indicated. See text for more details.
    The results are shown for unsmoothed field i.e. $n=-3$.
    We have used the parameterizations in Eq.(\ref{eq:fitting}) for various models. The Horndeski model is given by
    $\xi_{\kappa}=1, \xi_{\lambda}=0$ and the beyond Horndeski theories are given by $\xi_{\kappa}=1, \xi_{\lambda}=1$.
  For the GR we have $\xi_{\kappa}=0, \xi_{\lambda}=0$.}
  \label{fig:3Dnu}
\end{figure} 

The theoretical modelling of the bispectrum in modified gravity scenarios is more challenging than 
the power spectrum calculation. Typically a perturbative approach is adopted in the
quasilinear regime \citep{BB11}. In addition, a quasi-static approximation is used, i.e. metric perturbations
are varying slowly with time that they can be ignored.
Many extensions of the perturbative approach were considered in
the literature in recent years \citep{Bose1,Bose2}. Typically, this is achieved by introducing more freedom to
the kernels and validating or calibrating them using numerical simulations.
Indeed, others including variants of halo model predictions too have been proposed that can reproduce
the simulation results with varying degree of success.

In the literature, typically, two main  families of modified gravity theories are considered.
(A) models with Vainshtein-screening mechanism which includes the DGP model
  as well as the Horndeski \citep{Hordensky74} and beyond Horndeski theories
  \citep{BeyondHordensky1,BeyondHordensky2, BeyondHordensky3}
  and (B) the models with Chameleon-screening that includes the Hu-Sawicki $f({\rm R})$ model \citep{fofR}.
  In the DGP model \citep{DGP} the bispectrum from simulations can be reproduced using
  the GR expression by suitably modifying the power spectrum.
  The situation is somewhat more complicated for $f(R)$ theories.  The numerical modelling
  is more important at small scales where analytical results start to fail.
%
%
  %
  \subsection{Bernardeau \& Brax Models}
  Next, we first turn to different phenomenological toy models of modified gravity presented by \cite{BB11}.

  \noindent
      {\bf (a) Gamma $\gamma$ Model: }
  This model is generated by modifying the Euler equation of the Euler-Continuity-Poisson equation.
  In this model the gravitational field seen by massive particles (denoted as $\phi^{\rm eff}$) is different
  from the gravitational potential that solved the Poisson equation $\phi$. These two
  potentials are different and related by $\phi^{\rm eff} = (1+\epsilon)\phi$ through
  parameter $\epsilon(t)$ in the sub-horizon scale.
  
  In this parametrization the kernel $F_2$ in Eq.(\ref{eq:F2}) is modified to the following form:
  \ben
  && F_2(\bk_1,\bk_2) = {1\over 2}(1+\epsilon) +
  {1\over 2} {\bk_1\cdot\bk_2 \over k_1 k_2} \left ( {k_1\over k_2} + {k_2 \over k_1} \right )  +
  {1\over 2}(1-\epsilon)\left [ {\bk_1\cdot\bk_2 \over k_1 k_2} \right ]^2
  \label{eq:Brax}
  \een
  In general the parameter $\epsilon$ can be a function of scale factor $a$ or the wavelength $k$.
  For $\epsilon = {3/7}$ recover the expression given in Eq.(\ref{eq:F3}). The Lagrangian perturbation
  theory is often used to model quasilinear evolution of gravitational clustering. The Zel'dovich approximation
  is the linear order in Lagrangian perturbation theory. The bispectrum in the Zel'dovich approximation
  can be recovered from Eq.(\ref{eq:Brax}) $\epsilon=0$ \citep{Staro}. 
  \ben
  && \langle F_2 \rangle_{\rm 3D} = {{\epsilon + 2}\over \xred{3}}; \quad \langle F_2 \rangle_{\rm 2D} = {{\epsilon + 3}\over \xred{4}}.
  \een
  \noindent
  The actual value of the parameter $\epsilon$ can be computed using the linearised Euler-Continuity-Poisson equation,
  and assuming a parametric form for the growth fcator $f=  d\ln D_+/d\ln a \approx \Omega^\gamma_{\rm M}$. A convenient form
  of a fitting function can be obtained for values not too far from General Relativistic ($\Lambda$CDM) values.
  This model can be considered as a special case of Eq.(\ref{eq:fitting}) with $\kappa=1$ and $\epsilon = 1-{4/7}\lambda$.
  The smoothing includes a dependence on spectral index. In 2D we have 
  \ben
  && C_{21} = R_2 \left [ 4\langle F_2 \rangle_{\rm 2D} - {1\over 2} (n+2) \right ].
  \een
      {\bf (b) Beta ($\beta$) Model: }
  In the $\beta$ model proposed by \citep{BB11} where the expression for the kernel $F_2(\bk_1,\bk_2)$ we have:
  \ben
  && F_2(\bk_1,\bk_2)  = \left (  {3\nu_s \over 4} - {1\over 2}\right ) +
  {1\over 2} {\bk_1\cdot\bk_2 \over k_1 k_2} \left [ {k_1\over k_2} + {k_2\over k_1} \right ]
  +  \left (  {3  \over 2} - {3\nu_s \over 2}\right )  \left [ {\bk_1\cdot\bk_2 \over k_1 k_2} \right ]^2 
  \een
  where, the parameter $\nu_2$ can be related to the $\epsilon$ parameter in Eq.(\ref{eq:Brax})  $\epsilon = \xred{3 \over 2} \nu_s -2$.
  The parametric value for $\nu_2$  can be obtained in a manner similar to the $\gamma$ model. However, we would leave them
  unspeified. The angular average gives $\langle F_2\rangle_{\rm 3D} = {\nu_s/2}$ and similarly
  $\langle F_2\rangle_{\rm 2D} = (3\nu_s/2 +1)/4 $ for 2D
  and is independent of $t$. In these models the $\nu_2$ can in general be a function of $z$ as well as wave-number $k$.
  This model was also recently used in \citep{IB_MG} for computation of a related statistics
  known as integrated bispectrum. In general the parameter can also be a $k$ dependent parameter.
  The expression for $C_{21}$ has the following form:
  \ben
  && C_{21} =  R_2 \left [ {3 \over 2} \nu_s +1  -  {1 \over 2} (n+2) \right ].
  \een
  The power spectrum too gets modified due to changes in the kernel $F_2(\bk_1,\bk_2)$ at one-loop.
  The loop corrections to the linear power spectrum depends on the $F_2(\bk_1,\bk_2)$ and
  thus can also be used to constrain any departure from GR. 
%
\subsection{Horndeski and Beyond Horndeski in the  Perturbative Regime}
%
Horndeski theories are scalar-tensor theories with a single propagating degree of freedom
and are free from Ostrogradsky type instabilities. Horndeski theories 
The Horndeski theories have also been extended by considering what are also
known as the degenerate higher-order scalar-tensor (DHOST) theories.
The simplest extensions in the context of non-degenerate
scenarios are also known as the Galeyzes-Langlois-Piazza-Venizzi or GPLV theories \citep{BeyondHordensky1,BeyondHordensky2}.
The second-order kernel in these scenario
include a scale dependent additional term which changes the bispectrum \citep{Hirano}
that can be constrained using the staistics discussed here.
\bes
\ben
&& F_2(\bk_1,\bk_2,z) = \kappa_s(z) \alpha_s(\bk_1,\bk_2) - {2 \over 7}\lambda_s(z)\gamma(\bk_1,\bk_2); \\
&& \alpha_s(\bk_1,\bk_2) = 1 + {1 \over 2} (\bk_1\cdot\bk_2) {(k_1^2 + k_2^2) \over k_1^2 k_2^2};
\quad \gamma_s(\bk_1,\bk_2) = 1- {(\bk_1\cdot\bk_2)^2 \over k_1^2 k_2^2}.
\label{eq:define_alpha_beta}
\een
\ees
Taking angular averages we can see $\alpha_s(\bk_1,\bk_2) = 1$ and $\gamma(\bk_1,\bk_2)= 1/2$ which leads us
respectively in 2D to:
\ben
&& \langle F_2 \rangle_{\rm 2D} = \kappa_s(z)  - {1 \over 7} \lambda_s(z); \quad  
\langle F_2 \rangle_{\rm 3D} = \kappa_s(z)  - {4 \over 21} \lambda_s(z).
\label{eq:2Dand3D}
\een
Similar calculation in the Effective Field Theory (EFT) of dark energy framework can be found in \citep{CLV18}
\ben
&& C_{21} = 2\int_0^{r_s} dr D_+^4(z) {w^3(r) \over d_A^{4+2n}(r)}
\left [ \kappa_s(z) - {1 \over 7}\lambda_s(z) -{1\over 2}\kappa_s(z)(n+2)  \right ]
\bigg / \left ( \int_0^{r_s} d\,r D^2_+(z){ w^2(r) \over d_A^{2+n}(r)} \right )^2.
\label{eq:c21_k_lambda}
\een
In \cite{Marco_Cristomi} the following equivalent parameterization for the kernel $F_2$ was introduced:
\ben
&& F_2(\bk_1,\bk_3) = A_{\alpha}(z) \alpha_s(\bk_1,\bk_2) + A_{\gamma}(z)\gamma(\bk_1,\bk_2).
\een
In terms of the parameters $A_{\alpha}(z)$ and  $A_{\gamma}(z)$
\ben
&& C_{21} = 2\int_0^{r_s} dr D_+^4(z) {w^3(r) \over d_A^{4+2n}(r)}
\left [ A_\alpha(z) + {1 \over 2}A_\gamma(z) -{1\over 2} A_{\alpha}(z)(n+2)  \right ] 
\bigg / \left ( \int_0^{r_s} d\,r D_+^2(z){ w^2(r) \over d_A^{2+n}(r)} \right )^2.
\een
In general the parameters $A_{\alpha}(z)=\kappa_s(z)$, $A_{\gamma}(z)=-2/7\lambda(z)$ are time-dependent. 
For this model we have $\langle F_2 \rangle_{\rm 3D} = A_{\alpha} + {2\over 3} A_{\gamma} $ and  $\langle F_2 \rangle_{\rm 2D} = A_{\alpha} + {1\over 2} A_{\gamma}$.
It is important to notice that these theories have an important difference with GR and Horndeski theories.
The Horndeski theories are invariant under time-dependent spatial co-ordinate transformations.
The form for the $F_2$ kernels are fixed by existence of such symmetry. Many modified gravity theories
fall under this category. In Beyond Horndeski theories, the fluid equations and the equations of gravity
possess very different symmetry properties and have a kernel $F_2$ that is structurally different.
This is related to violation of these theories from the so-called consistency relation
which are respected in GR \citep{consistency}. Future surveys such as the {\em Euclid} survey will be able to
probe such theories beyond the consistency relations using the statistics developed here.

A detailed study for the skew-spectrum \citep{MunshiHeavens} Minkowski functionals\citep{waerbeke} for these models and the integrated bispectrum \citep{IB}
as well as the related consistency relations will be presented elsewhere (Munshi et al. 2020, in preparation).

\subsection{Normal-branch of Dvali, Gabadadze, Porrati (nDGP) model}
The normal branch of \citep{DGP} model known also as the nDGP is a
    prototypical example that involve Vainshtein screening.
The model of bispectrum that is known to accurately reproduce the bispectrum was computed by \citep{KTH} which
correspond to the case $\kappa=1$ in Eq.(\ref{eq:fitting}).
\ben
&& \kappa_s(z)=1 ; \quad\quad \lambda_s(z)= \left ( 1- {7\over 2} {D_2(z) \over D^2_+(z)}\right ).
\een
Here $D_2(z)$ and $D_+(z)$ are the first-order and second-order growth factors
that can be computed by numerically solving the equations governing growth of perturbations \citep{Bose1}.
\begin{figure}
  \centering
  \includegraphics[width=6cm]{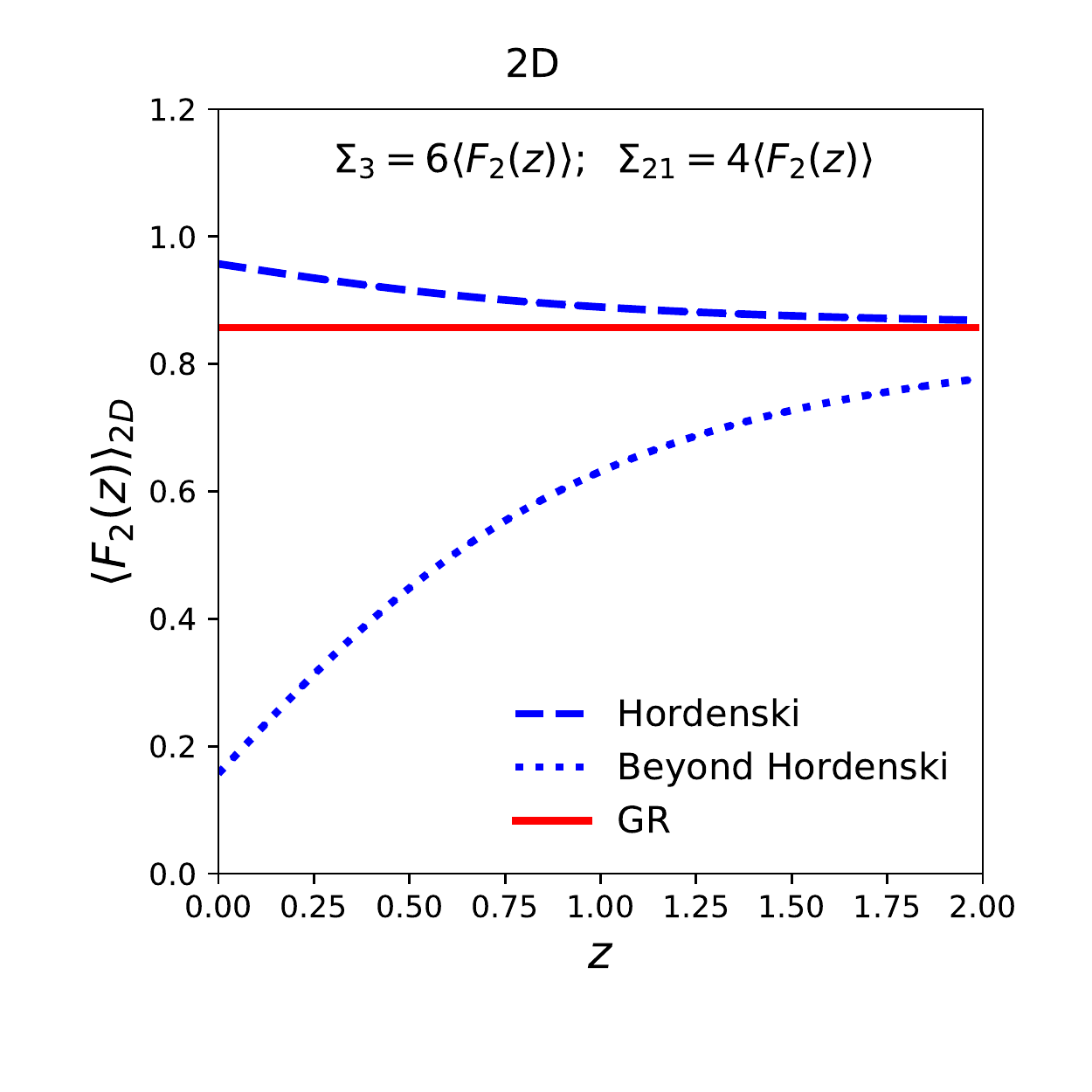}
  \includegraphics[width=6cm]{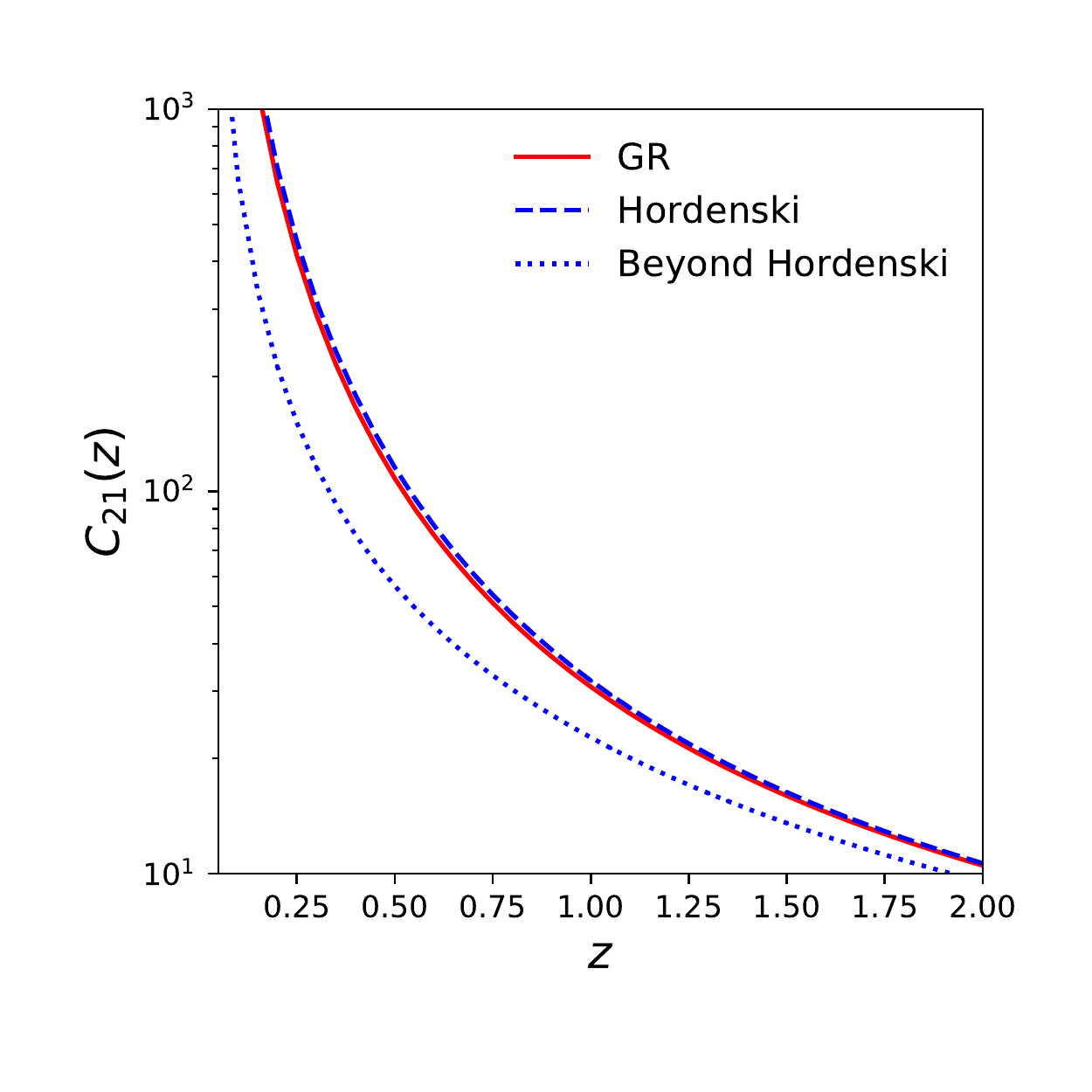}
  \caption{The left panel shows $\langle F_2(z) \rangle_{\rm 2D}$ for GR, Horndeski and Beyond Horndeski as a function
    of redshift $z$ in 2D. The results are plotted for $n=-2$ which represents the unsmoothed field.
    The right panel corresponds to $C_{21}(z)$ as a function for these models.}
  \label{fig:2Dnu}
\end{figure} 
%
\subsection{Massive Neutrinos}
%
A small but non-negligible fraction of the cosmological matter density is
provided by massive neutrinos \citep{nu}. The massive neutrinos are known to have
significant thermal distribution and a different cosmological
evolutionary history in comparison to the cold dark matter.
The thermal dispersion in velocity results in a damping of perturbation
below a length scale also known as the free-streaming length scale.
This will be probed by future surveys with a very high degree of accuracy.
In the long run cosmological surveys are expected to provide
an upper limit to the sum of the neutrino masses.. This will be very
useful when jointly considered with the lower limits from
neutrino-oscillation experiments.

The neutrinos decouple and free-stream with a large thermal velocities.
The redshift $z_{nr}$ at which neutrinos become relativistic
depend on their mass eigenstate $m_i$,  $1+ z_{nr} = 1980 \left [ {m_{\nu,i}/1{\rm eV} }\right ]$
The fractional contribution to the total matter density is denoted as $f_{\nu}$
which can be expressed as
\ben
&& f_{\nu} \equiv {\Omega_{\nu} \over \Omega_{M}} = {1 \over \Omega_{M,0} h^2} {\sum_i M_{\nu,i} \over 93.14 \rm eV}.
\een
In future it will also be interesting to consider the effect of neutrino mass
on bispectrum when simulated all-sky lensing maps for such cosmologies will be available \citep{Liu17,Coulton18}.
The total matter distribution thus can be written in terms of the cold dark matter perturbation $\delta_{cdm}$
and the fluctuations in the neutrino density distribution $\delta_{\nu}$.
\ben
&& \delta_m = f_c\delta_c + f_{\nu}\delta_{\nu}; \quad \ f_c+f_\nu=1.
\een
The resulting matter power spectrum $P_{mm}(k)$ and bispectrum $B_{mmm}(\bk_1,\bk_2,\bk_3)$ can be expressed as \citep{Ruggeri}:
\bes
\ben
&& P_{mm}(k) = f_c^2 P_{cc}(k) + 2f_{\nu}f_c P_{\nu c}(k) + f^2_{\nu} P_{\nu\nu}(k)\\
&& B_{mmm} = f_c^3 B_{ccc} + f_c^2 f_{\nu} B_{cc\nu} + f_c f^2_{\nu} B_{\nu\nu c} + f^3_{\nu} B_{\nu\nu\nu}.
\een
\ees
Here $P_{cc}$ and $P_{\nu\nu}$ represent the power spectrum cold dark matter and the neutrino component where as the $P_{\nu c}$
is the cross spectra between them. We will drop the suffix 3D to avoid cluttering. 
We will only consider the linear order perturbation in $\delta_\nu$ and ignore all higher order contributions which implies $B_{\nu\nu\nu} =0.$
For  $B_{ccc}$ the expression in the squeezed limit is exactly same as derived before.
\bes
\ben
&& B^{\rm 2D, sq}_{ccc} = \left [ {24 \over 7} -{1 \over 2}{ d k^2 \,P_{cc}(k) \over d\ln k  } \right ]P_{cc}(k_\perp)P_{cc}(q_{3\perp}). 
\een
\ees

We will next consider the mixed terms $B_{\nu\nu c}$ 
These contributions in terms of $\delta_c$ and $\delta_{\nu}$ can be expressed as:
\bes
\ben
&& B_{cc\nu}(\bk_1,\bk_2,\bk_3) =  \langle\delta_c(\bk_1)\delta_{c}(\bk_2)\delta_{\nu}(\bk_3)\rangle +  {\rm cyc.perm.}; \\
&& B_{\nu\nu c}(\bk_1,\bk_2,\bk_3) =  \langle\delta_\nu(\bk_1)\delta_{\nu}(\bk_2)\delta_{c}(\bk_3)\rangle. + {\rm cyc.perm.}. 
\een
\ees
In the above equations the cyc. perm. represent  cyclic permutations of the wave vectors $\bk_1,\bk_2$ and $\bk_3$.

To evaluate $B_{\nu\nu c}$ we expand the terms  perturbatively. Employing tree level perturbation theory, the contributions from  $B_{\nu\nu c, 112}$ are
from these terms:
\ben
&& B_{\nu\nu c} =  B_{\nu\nu c, 112}(\bk_1,\bk_2,\bk_3) +  B_{\nu\nu c, 112}(\bk_2,\bk_3,\bk_1) + B_{\nu\nu c, 112}(\bk_1,\bk_2,\bk_3).
\een
In our notation, $B_{\nu\nu c, 112}(\bk_1,\bk_2,\bk_3) \equiv \langle \delta_\nu^{(1)}(\bk_1)\delta_\nu^{(1)}(\bk_2)\delta_c^{(2)}(\bk_3)\rangle$
and similarly for the other terms. In tems of the second-order kernels $F_2(\bk_1,\bk_2)$ we have:
\ben
&& B_{\nu\nu c,112}(\bk_1,\bk_2,\bk_3)= 2F_2(\bk_1,\bk_2) P_{\nu c}(k_1) P_{\nu c}(k_2)  \, ; 
\een
The other terms can be recovered by cyclic permutation of the wavenumber. In the squeezed limit we have:
\ben
&& B^{\rm 2D, sq}_{\nu\nu c} = \left [ {24 \over 7} -{1 \over 2}{ d \ln k^2 \,P_{\nu c}(k_\perp) \over d\ln k_\perp  } \right ]P_{\nu c}(k_\perp)P_{\nu c}(q_{3\perp}). 
\een
Finally we turn to $B_{cc \nu}$. The perturbative contributions are as follows:
\ben
 && B^{\rm 2D, sq}_{\nu\nu c} = 2[ F_2(\bk_1,\bk_2) P_{cc}(k_1)P_{c\nu}(k_2) + {\rm cyc.perm.} ].
 \een
Going through an elaborate algebraic manipulation we arrive at the squeezed limi:
\ben
&& B^{\rm 2D, sq}_{\nu\nu c} = \left [ {24 \over 7} -{1\over 2} {d \ln k^2 P_{cc}(k_\perp) \over d \ln k_\perp }
  \right ]P_{cc}(k_\perp)P_{cc}(q_{3\perp}) 
+ \left [ {24 \over 7} -{1\over 2} {d k_\perp^2 P_{c\nu}(k_\perp) \over d \ln k_\perp } \right ]
P_{c\nu}(k_\perp)P_{cc}(q_{3\perp}).
\een
%
In future it will be interesting to study the effect of neutrino mass
on bispectrum using simulations when  all-sky lensing maps for such cosmologies will be available \citep{Liu17,Coulton18}.
%
\subsection{Clustering Quintessence}
%
%
Quintessence \citep{Q_review}
is the most popular dynamics of dark energy in which the
potential energy of a single scalar field drives the accelerated
expansion of the Universe.  The quintessence model is different from
the cosmological constant scenario allowing for a different temporal dependence of the observables.
The scalar field in
most quintessence models is considered homogeneous and
is typically minimally coupled. The sound speed of the scalar field
in these models equals the speed of light which prevents
any clustering below the horizon scale. However, extensions
of such models with vanishing or lower than speed of light
have also been considered. These models are known as the clustering quintessence models\citep{Sefusatti, Spherical_DE}.
The future large scale structure surveys
can be used to differentiate between these two scenarios. We use our formalism to derive the changes in
the bispectrum in the squeezed limit in these models.
We quote the expression of the kernel $F_2$ from \citep{Sefusatti}:
\bes
\ben 
&& {D_+ \over a} = {5 \over 2} \Omega_{M} \left [ {\Omega_{M}}^{4/7} + {3 \over 2} \Omega_M
  + \left ({1 \over 70}- {1+w \over 4} \right ) \Omega_Q \left ( 1 + {\Omega_{M} \over 2} \right )\right ]^{-1}.
\een
\ees
Here, $\Omega_{\rm Q}$ and $\Omega_{\rm M}$ are the density parameter related to Quintessence and dark matter.
The corresponding linear growth rates are denoted by $D_{Q+}$ and $D_+$.
The parameters $\epsilon_s = {\Omega_Q \over \Omega_M} {D_{Q,+} \over D_{+}}$ and $\nu_2$
can also be expressed in terms of $\Omega_{\rm Q}$ and  $\Omega_{\rm M}$ and depend of redshift $z$.
\bes
\ben
&& F_2(\bk_1,\bk_2,\eta) = {\nu_s \over 2} + {1 \over 2}(1-\epsilon_s){ \bk_1\cdot \bk_2 \over k_1 k_2} \left ( {k_1 \over k_2} + {k_2 \over k_1} \right )
- {1 \over 2}\left ( 1-\epsilon_s -{\nu_s \over 2} \right ) \left [ 1- 3 \left ( {\bk_1\cdot \bk_2 \over k_1 k_2} \right )^2 \right ].
\een
Thus, two different parameters $\epsilon_s(z)$ and $\nu_s(z)$ to describe the tree-level bispectrum in this model.
\ees
 \bes
 \ben
 && C_{21} = \int_0^{r_s} dr {w^3(r) \over d^{4+2n}(r)} D_+^4(z)
  \Big [ {1 \over 4}  (1-\epsilon_s) + {3\over 8}\nu_s -{1\over 2}(1-\epsilon_s)(n+2)  \Big ]
 \Bigg / \left( \int_0^{r_s} dr {w^2(r) \over d^{2+n}(r)} D_+^2(z) \right )^2.
\een
\ees
Typically at low redshift for some values of $w$ the parameter $\epsilon$ can reach upto $10\%$ which
can lead to roughly an order of $10\%$ correction to the bispectrum which can be accounted for
high precision measurements from future surveys.
%
\subsection{Bispectrum in General Scalar-tensor Theories}
  %
Next, we consider a phenomenological fitting function.
  The second-order perturbative analysis of the general scalar tensor theories were initially performed by \citep{Hirano}
  which was later extended to smaller non-perturbative scales using a fitting function \citep{Bose2},
  Using the fitting function proposed in \citep{Bose2} we can compute the $C_{21}$ in a class of models which
are represented by the following expression for $F_2(\bk_1,\bk_2,z)$ replacing $F_2(\bk_1,\bk_2)$ in Eq.(\ref{eq:F3}):
\bes
\ben
&& F_2(\bk_1,\bk_2,z) = \left [ \kappa_s(z) - {2 \over 7}\lambda_s(z) \right ]a(k_1,z)a(k_2,z) 
+ {1\over 2} \kappa_s(z)  \left [ {\bk_1\cdot\bk_2 \over k_1 k_2}  \right ]\left ( {k_1\over k_2}+ {k_2\over k_1}\right )
b(k_1,z)b(k_2,z) \nn \\
&& \hspace{2cm} + {2\over 7} \lambda(z)  \left [ {\bk_1\cdot\bk_2 \over k_1 k_2}  \right ]^2 c(k_1,z)c(k_2,z);  \label{eq:fitting}\\
&& \lambda_s(z) = [\Omega_{\rm M}(z)]^{\xi_{\lambda}} ;  \quad  \kappa_s(z) = [\Omega_{\rm M}(z)]^{\xi_{\kappa}}; \quad
\Omega_{\rm M}(z) = \Omega_{\rm M,0}(1+z)^3 /((1+z)^3\Omega_{\rm M,0} + \Omega_{\Lambda}).
\label{eq:fitting}
\een
\ees
The functions $\kappa_s(z)$ and $\lambda(z)$ are approximated using the above functional forms and $\xi_{\lambda}$ and 
are free parameters that can be estimated from observational data. The functional forms for $a, b$ and $c$ are assumed to be
same as that of their $\Lambda$CDM form \citep{SC01,GM11} which interpolates the perturbative regime and highly nonlinear
assumed to be described by Hyper-Extended-Perturbation-Theory \citep{Scoccimarro}.
To be consistent with the literature we have used $\kappa_s$ to denote one of the parameters which should not be
confused with the weak lensing convergence $\kappa$ as their meaning would be obvious from the context. 
For $\kappa_s(a) =1$ and $\lambda_s(a)=1$ or, equivalently, $\xi_{\kappa}=0$ and $\xi_{\lambda}=0$ we
recover the case of General Relativity (GR) presented in Eq.(\ref{eq:F3}).
As discussed before, the Horndeski theories are the most general class of scalar-tensor theories which are non-degenerate that
leads second-order equations of motion in 4D. In these models, $\lambda_s\ne 1$ though $\kappa = 1$ still remains valid.
A generalization of Horndeski \citep{Hordensky74} theory leads to a class of models that are known as
``Beyond Horndeski'' models \citep{BeyondHordensky1,BeyondHordensky2,BeyondHordensky3}.
In his models both $\kappa_s$ and $\lambda_s$ can deviate from unity. At high-$z$ the both theories converge to GR
as expected. The Horndeski theories violate the Vaishentein mechanism to recover GR at nonlinear
scale has also been considered. In these scenarios the parameter both $\kappa$ and $\lambda$ deviates from
unity. Thus testing GR which correspond to $\lambda=\kappa=1$ reduces to constrain
deviation of $\lambda$ and $\kappa$ from unity. The functional form for $\kappa$ and $\lambda$ is adopted
from \citep{Namikawa18} and converges to GR at high-z as expected.

We will next focus on computing the second order vertex $\nu_2$ as defined in Eq.(\ref{eq:define_nuN}) for both 3D and 3D.
Unlike in case of GR, in general these vertices have a redshift dependence. To compute these we start by noticing that
in both three and two dimensions we have $\left \langle {\bk_1\cdot\bk_2 / k_1 k_2}  \right \rangle=0$ and in
2D we have $\left\langle \left ({\bk_1\cdot\bk_2/k_1 k_2}\right )^2 \right\rangle = {1/2}$. In the following,
we will ignore smoothing as the correction terms involved will be exactly same as the one presented 

In the quasilinear regime the functions $a,b$ and $c$ tend to unity. In this limiting case
the departure from GR is encoded only in the redshift dependent factors and the expression for
$C_{21}$ is identical to Eq.(\ref{eq:c21_k_lambda}) with the  specific form for $\kappa_s$ and $\lambda_s$
are given by Eq.(\ref{eq:fitting}).
Substituting $\kappa_s(z)=1$ and  $\lambda_s(z)=1$ we recover the unsmoothed results for GR.
The smoothing in 3D and 2D will introduce terms involving factors of $(n+3)$ in Eq.(\ref{eq:3DS3})-Eq.(\ref{eq:3DC21})
and $(n+2)$ in Eq.(\ref{eq:2DS3})-Eq.(\ref{eq:ib_def}).
The results for specific models for 3D and 2D are respectively shown in Figure-\ref{fig:3Dnu} and Figure-\ref{fig:2Dnu}.
While for GR the $\langle F_2 \rangle$ is
independent of redshift $z$, for Horndeski and beyond Horndeski theories $\langle F_2 \rangle$ depends on redshift.
At higher $z$ they become identical to that of GR as expected. In Figure-\ref{fig:3Dnu} the $\langle F_2 \rangle$ for
the 2D cylindrical collapse is plotted as function of $z$ and their
pattern of evolution is same as in 3D. The effect of line-of-sight projection is encoded in
the factor $R_2(z)$ which is shown in the right panel.

Although, the results for higher-order spectra are known to an arbitrary order in GR, similar
results for most of the modified gravity theories are known mostly to second order.
Going beyond third-order in general requires order-by-order calculation.
While we have considered the statistics of 3D density
field $\delta$ and resulting convergence $\kappa$ similar results can be obtained
for the divergence of peculiar velocity.

The tests involving bispectrum related statistics presenetd here 
can further tighten the constraints obtained using linear growth rate alone.
This is particularly important as no strong constraint on $\lambda_s$ and $\kappa_s$ exist currently.
Indeed, there are no upper or lower limits for $\kappa_s$ based on theoretical expectation.
  
Before we conclude this section, we would like to point out that the two paramters used in defining the clusteing
quintessence i.e. $\nu_s$ and $\epsilon_s$ (or $\alpha_s$ and $\beta_s$ for the case of DHOST theories)
can independently be contrained using 3D and 2D measuremenets. This is due to the fact that
the statistics $C_{21}$ depends on $\nu_s$ and $\epsilon$ in a different manner in 3D and 2D.
We have concentrated on projected or 2D surveys in this paper but similar results will be presented for 3D surveys
in a separate article. 
%
\section{Conclusions and Future Prospects}
\label{sec:conclu}
%
We have computed the skew-spectrum (see Eq.(\ref{eq:defSkew})) and kurtosis-spectrum  Eq.(\ref{eq:defK31}))
at low $\ell$ for the analysis of weak lensing convergence or $\kappa$ maps. These spectra
generalizes the one-point cumulants,
e.g. the skewness  and kurtosis defined in \textsection{\ref{sec:skew}, and
are often used in the literature for
analyzing higher-order non-Gaussianity of cosmological maps. They capture some of the essential properties of the full bispectrum
or trispectrum which are more difficult to estimate.
In the real space these spectra correspond to cumulant correlators that can be computed in the leading-order using tree-level perturbations
in the large-separation limit. In this limit these spectra can be computed to arbitrary order using tree-level perturbative calculations without any
need for any phenomenological fitting functions or extensions of perturbative calculation.
We use the flat-sky approximation and Eulerian perturbative results based on generating function approach we show how to compute
high-order spectra to arbitrary order. We test these results for lower-order spectra namely the skew- and kurt-spectra against state-of-the-art
all-sky weak lensing maps. We find the our results are in good agreement. These results will be valuable in analyzing higher-order statistics
from future all-sky weak lensing surveys such as the {\em Euclid} survey
The presence of mask generated from near all-sky surveys introduces mode mixing. Unless corrected, the mode mixing introduced by a
mask can be a source of confusion while analyzing the higher-order spectra as they encode information about gravity induced mode-coupling
We have presented a generalization of existing method typically used in the study of ordinary power spectrum to construct an unbiased
estimates of higher-order spectra Eq.(\ref{eq:MLL1})-Eq.(\ref{eq:MLL2}).

The parameters $C_{pq}$ computed for 3D weak lensing will be important when photo-z information is available. The statistics introduced
here will be useful in analyzing non-Gaussianity in such context. We will present results of such analysis in future work.
The results presented here can be generalized using a 3D Fourier-Bessel transform or a 3D flat sky formalism.
As noted before the 3D analysis allows factorization $C_{pq}=C_{pq}C_{q1}$ and their dependence on the spectral index $n$
are different so 2D and 3D results will provide independent information as well as much needed consistency checks and test for
possible systematics.

Any modification of gravity leaves detectable signature at the level of bispectrum.
Though such signatures are less prominent than any modification at the level of power spectrum,
it has recently attracted a lot of attention in the context of CMB lensing bispectrum \citep{Namikawa18}.
Similar investigations in the context of weak lensing are currently being pursued using
various statistical estimators. Various techniques were adopted
to extend perturbaive results derived in the context of General Relativity (GR).
Extensions to modified gravity scenarios were implemented by introducing more freedom to the kernels and calibrating then using
numerical simulations \citep{Bose1}. The expressions for bispectrum
exist for both type of modified gravity scenarios i.e. 
models with Vainshtein-screening mechanism which includes the DGP model as well as the Horndeski \citep{Hordensky74}
and  beyond Horndeski theories \citep{BeyondHordensky1,BeyondHordensky2, BeyondHordensky3}.
In the other class of models i.e. models with  Chameleon-screening that includes the Hu-Sawicki $f({\rm R})$ model \citep{fofR}
the bispectrum from simulations can be successfully reproduced using the GR expression
but with suitable modification of the power spectrum. We will extend our results
derived here to the modifying gravity scenarios as well as scenarios involving
massive neutrinos.

The position-dependent bispectrum and its higher-order generalization at the level of trispectrum
has exact one-to-one correspondence with the statistics studied in this paper. Indeed the
expressions for integrated bispectrum and the skew-spectrum at low-$\ell$ are identical.
However, the physical interpretation is different. The expressions at the level of fourth order
are not the same. The integrated bispectrum or equivalently the position-dependent power spectrum
probes the influence of large scale modes on small-scale structure. The cumulant correlators
at large separation limit as well as their harmonic counterparts namely the skew-spectrum and
kurt-spectra probe dynamics mainly at scales of smoothing. Comparing results from
these two statistics can provide useful cross-checks at each order.
 
Finite sky coverage can introduce bias in our estimators. The scatter and bias introduced
by finite survey size have been studied in great detail for galaxy surveys and to a lesser
extent for weak lensing surveys \citep{MuCo13}. These are less dominant in the quasi-linear regime
where the variance is small in the limiting case which we have studied here.

In our study we have assumed that the bispectrum is of even parity. Many studies in the recent past
have pointed out existence of an off-parity
bispectrum \citep{cmb_minkowski}. Such a bispectrum do not arise from 3D density perturbations.
However, signatures of contributions can be used to test possible existence of systematics.

In a recent work \citep{gameof19}, it was shown that {\em nulling} can be used effectively
to improve the accuracy of perturbative calculations by reducing the cross-talk between
quasilinear and nonlinear scales. These calculations were performed in the real-space
focusing primarily on one-point cumulants and PDF. In contrast our results here
concern primarily on two-point correlators and their associated spectra in the Fourier domain.
Applying the nulling before computing the spectra is expected to improve the validity
of the perturbative results.

Last but not least, the next generation of CMB Stage-IV experiments will be able to map
the projected lensing potential all the way to the surface of last scattering. It is expected
that the results obtained in this paper will be valuable in analyzing higher-order statistics
of maps obtained from such experiments \citep{Aba}. However, in this case the estimator described
here will have to be optimized to tackle low signal-to-noise for higher-order statistics of CMBR.
The post-born corrections \citep{LewisPratten}  play an important role in higher-order statistics of CMBR.
For realistic comparison against observations such corrections should be included. 
%
%
\section*{Acknowledgment}
DM is supported by a grant from the Leverhume Trust at MSSL.
It is a pleasure for DM to thank F. Bouchet, T. D. Kitching, T. Namikawa,   R. Takahashi, A. Taruya  and F. Vernizzi for many useful discussions.
We would like to also thank R. Takahashi for making the lensing maps publicly available.
We would like to also thank R. Schoenrich for careful reading of the draft and many suggestions that greatly improved the presentation.
DM would also like to organizers of the Euclid Theory Working Group Meeting (8th April - 9th, April 2019) in Oxford.
%
\bibliography{skew.bbl}
%
\end{document}

%% file: local-macro.tex
\def\la{\langle}
\def\ra{\rangle}

\def\be{\begin{equation}}
\def\ee{\end{equation}}
\def\ben{\begin{eqnarray}}
\def\een{\end{eqnarray}}
\def\nn{\nonumber}
\def\oh{\bf \hat\Omega}

\def\bk{{\bf k}}

\def\br{{\bf r}}

\def\bk{{\bf k}}

\def\bl{{\bf l}}

\def\2p{{(2\pi)^2}}

\def\bl{{\bf l}}
\def\be{\begin{equation}}
\def\ee{\end{equation}}
\def\beq{\begin{equation}}
\def\eeq{\end{equation}}
\def\ben{\begin{eqnarray}}
\def\een{\end{eqnarray}}
\def\bes{\begin{subequations}}
\def\ees{\end{subequations}}

\def\oh{{\hat\Omega}}
\def\nn{{\nonumber}}

\newcommand{\beqa}{\begin{eqnarray}}
\newcommand{\eeqa}{\end{eqnarray}}

\newcommand{\rA}{{\rm A}}

\def\ikap0{{\cal J}_{\theta_0}(r)}

\def\one1{\langle \kappa_{(i)}\kappa_{(j)} \rangle}
\def\one{{[\bar \xi^{(ij)}]}}

\def\ba{\begin{eqnarray}}
\def\ea{\end{eqnarray}}

\def\bk{{\bf k}}

\def\br{{\bf r}}

\def\bk{{\bf k}}

\def\bl{{\bf l}}

\def\2p{{(2\pi)^2}}

\def\bl{{\bf l}}

\def\be{\begin{equation}}
\def\ee{\end{equation}}

\def\beq{\begin{equation}}
\def\eeq{\end{equation}}

\def\ben{\begin{eqnarray}}
\def\een{\end{eqnarray}}

\def\oh{{\hat\Omega}}
\def\nn{{\nonumber}}

\def\bk{{\bf k}}

\def\bl{{\bf l}}

\def\2p{{(2\pi)^2}}

\def\bl{{\bf l}}

\def\bl{{\bf{l}}}

\def\btheta{{\boldsymbol\theta}}
\def\oh{{\bf{\widehat\Omega}}}

\def\rA{{\rm{A}}}
\def\rB{{\rm{B}}}